# Modeling share prices of banks and bankrupts

Ivan O. Kitov, IDG RAS

**Abstract**
Share prices of financial companies from the S&P 500 list have been modeled by a linear function of consumer price indices in the USA. The Johansen and Engle-Granger tests for cointegration both demonstrated the presence of an equilibrium long-term relation between observed and predicted time series. Econometrically, the pricing concept is valid. For several companies, share prices are defined only by CPI readings in the past. Therefore, our empirical pricing model is a deterministic one. For a few companies, including Lehman Brothers, AIG, Freddie Mac and Fannie Mae, negative share prices could be foreseen in May-September 2008. One might interpret the negative share prices as a sign of approaching bankruptcies.

## Introduction

Recently, we have developed and tested statistically and econometrically a deterministic model predicting share prices of selected S&P 500 companies (Kitov, 2010). We have found that there exists a linear link between various subcategories of consumer price index (CPI) and some share prices, with the latter lagging by several months. In order to build a reliable quantitative model from this link one needs to use standard and simple statistical procedures.

Following the general concept and principal results of the previous study, here we are predicting stock prices of financial companies from the S&P 500 list. In several cases, robust predictions are obtained at a time horizon of several months. In close relation to these financial companies we have also investigated several cases of bankruptcy and bailout. These cases include Lehman Brothers (LH), American International Group (AIG), Fannie Mae (FNM) and Freddie Mac (FRE). Regarding these bankruptcies, we have tested our model against its predictive power in May and September 2008. The main question was: Could the bankruptcies be foreseen? If yes, which companies should or should not be bailed out as related to the size of their debt?

In the mainstream economics and finances stock prices are treated as not predictable beyond their stochastic properties. The existence of a deterministic model would undermine the fundamental assumption of the stock market. If the prices are predictable, the participants would have not been actively defining new prices in myriads of tries, but blindly followed the driving force behind the market. It is more comfortable to presume that all available information is already counted in. However, our study has demonstrated that the stochastic market does not mean an unpredictable one.

In this paper, we analyze sixty six financial companies from the S&P 500 lists as of January 2010 as well as a few bankrupts from the financials. Some of the companies have been accurately described by models including two CPI subcategories leading relevant share prices by several months. Other companies are characterized by models with at least one of defining CPI components lagging behind related stock prices. We have intentionally constrained our investigation to S&P 500 - we expect other companies to be described by similar models.

Our deterministic model for the evolution of stock prices is based on a "mechanical" dependence on the CPI. Under our framework, the term "mechanical" has multiple meanings. Firstly, it expresses mechanistic character of the link when any change in the CPI is one-to-one converted into the change in related stock prices, as one would expect with blocks or leverages. Secondly, the link does not depend on human beings in sense of their rational or irrational behavior or expectations. In its ultimate form, the macroeconomic concept behind the stock price model relates the market prices to populations or the numbers of people in various age groups irrelevant to their skills. Accordingly, the populations consist of the simplest possible objects; only their numbers matter. Thirdly, the link is a linear one, i.e. the one often met in classical mechanics. In all these regards, we consider the model as a mechanical one and thus a physical one rather than an economic or financial one. Essentially, we work with measured numbers not with the piles of information behind any stock.

For the selected stocks, the model quantitatively foresees at a several month horizon. Therefore, there exist two or more CPI components unambiguously defining share prices several months ahead. It is worth noting that the evolution of all CPI components is likely to be defined, in part, by stochastic forces. According to the mechanical dependence between the share prices and the CPI, all stochastic features are one-to-one converted into stochastic behavior of share prices. Since the prices lag behind the CPI, this stochastic behavior is fully predetermined. The predictability of a measured variable using independent measured variables, as described by mathematical relationships, is one of the principal requirements for a science to join the club of hard sciences. Therefore, our stock pricing model indicates that the stock market is likely an object of a hard science.

A model predicting stock prices in a deterministic way is a sensitive issue. It seems unfair to give advantages to randomly selected market participants. As thoroughly discussed in (Kitov, 2009b; Kitov and Kitov, 2008; 2009ab) the models are piecewise ones. A given set of empirical coefficients holds until the trend in the difference between defining CPI is sustained. Such sustainable trends are observed in a majority of CPI differences and usually last between 5 and 20 years (Kitov and Kitov, 2008). The most recent trend has been reaching its natural end since 2008 and the transition to a new trend in 2009 and 2010 is likely the best time to present our model. As a result, there is no gain from the empirical models discussed in this paper. Their predictive power has been fading away since 2008. When the new trend in the CPI is established, one will be able to estimate new empirical coefficients, all participants having equal chances.

The remainder of the paper is arranged as follows. Section 1 introduces the model and data, which include stock prices of sixty six S&P 500 financial companies and seventy CPI components. In Section 2, empirical models are presented both in tabulated and graphical forms. For each model we have estimated standard deviation, which serves as a proxy to the model accuracy. For a few companies, the estimated models are robust over the previous 10 months. Section 3 tests these models statistically and econometrically. The Johansen (1988) and Engle-Granger (Newbold and Granger, 1967; Hendry and, Juselius, 2001) tests both demonstrate that the null hypothesis of the existence a cointegrating relation between the observed and predicted time series cannot be rejected for a majority of companies. Therefore, the model is justified econometrically, and thus, all statistical inferences are valid. In Section 4, a crucial historical problem is addressed: Could one predict in May 2008 the evolution of financial stock prices? For some companies, the models estimated in the beginning of 2008 hold over the next year. Hence, the empirical modeling would have allowed accurate prediction of the evolution of stock prices, including those related to companies who filed for bankruptcy in several months. Finally, Section 5 investigates several cases of bankruptcy and bailout in the United States. It is found that many stock price trajectories would have been predicted to dive below the zero line.

The results of the presented research open a new field for the future investigations of the stock market. We do not consider the concept and empirical models as accurate enough or final. There should be numerous opportunities to amend and elaborate the model. Apparently, one can include new and improve available estimates of consumer price indices.

### 1. Model and data

Kitov (2009b) introduced a simple deterministic pricing model. Originally, it was based on an assumption that there exists a linear link between a share price (here only the stock market in the United States is considered) and the differences between various expenditure subcategories of the headline CPI. The intuition behind the model was simple - a higher relative rate of price growth (fall) in a given subcategory of goods and services is likely to result in a faster increase (decrease) in stock prices of related companies. In the first approximation, the deviation between price-defining indices is proportional to the ratio of their pricing powers. The presence of sustainable (linear or nonlinear) trends in the differences, as described in (Kitov and Kitov, 2008; 2009ab), allows predicting the evolution of the differences, and thus, the deviation between prices of corresponding goods and services. The trends are the basis of a long-term prediction of share prices. In the short-run, deterministic forecasting is possible only in the case when a given price lags behind defining CPI components.

In its general form, the pricing model is as follows (Kitov, 2010):

$$sp(t_j) = \Sigma b_i \cdot CPI_i(t_j - \tau_i) + c \cdot (t_j - 2000) + d + e_j \quad (1)$$

where $sp(t_j)$ is the share price at discrete (calendar) times $t_j$, $j=1,\ldots,J$; $CPI_i(t_j-\tau_i)$ is the $i$-th component of the CPI with the time lag $\tau_i$, $i=1,\ldots,I$; $b_i$, $c$ and $d$ are empirical coefficients of the linear and constant term; $e_j$ is the residual error, which statistical properties have to be scrutinized. By definition, the bets-fit model minimizes the RMS residual error. The time lags are expected because of the delay between the change in one price (stock or goods and services) and the reaction of related prices. It is a fundamental feature of the model that the lags in (1) may be both negative and positive. In this study, we limit the largest lag to fourteen months. Apparently, this is an artificial limitation and might be changed in a more elaborated model. In any case, a fourteen-month lag seems to be long enough for a price signal to pass through.

System (1) contains $J$ equations for $I+2$ coefficients. Since the sustainable trends last more than five years, the share price time series have more than 60 points. For the current recent trend, the involved series are between 70 and 90 readings. Due to the negative effects of a larger set of defining CPI components discussed by Kitov (2010), their number for all models is ($I=$) 2. To resolve the system, we use standard methods of matrix inversion. As a rule, solutions of (1) are stable with all coefficients far from zero.

At the initial stage of our investigation, we do not constrain the set of CPI components in number or/and content. Kitov (2010) used only 34 components selected from the full set provided by the US Bureau of Labor Statistics (2010). To some extent, the original choice was random with many components to be similar. For example, we included the index of food and beverages and the index for food without beverages. When the model resolution was low, defining CPI components were swapping between neighbors.

For the sake of completeness we always retain all principal subcategories of goods and services. Among them are the headline CPI ($C$), the core CPI, i.e. the headline CPI less food and energy ($CC$), the index of food and beverages ($F$), housing ($H$), apparel ($A$), transportation ($T$), medical care ($M$), recreation ($R$), education and communication ($EC$), and other goods and services ($O$). The involved CPI components are listed in Appendix 1. They are not seasonally adjusted indices and were retrieved from the database provided by the Bureau of Labor Statistics (2010). Many indices were started as late as 1998. It was natural to limit our modeling to the period between 2000 and 2010, i.e. to the current long-term trend.

Since the number and diversity of CPI subcategories is a crucial parameter, we have extended the set defining components to 70 from the previous set of 34 components. As demonstrated below, the extended set has provided a significant improvement in the model resolution and accuracy. Therefore, we envisage the increase in the number and diversity of defining subcategories as a powerful tool for obtaining consistent models. In an ideal situation, any stock should find its genuine pair of CPI components. However, the usage of similar components may have a negative effect on the model – one may fail to distinguish between very close models.

Every sector in the S&P 500 list might give good examples of companies with defining CPI components lagging behind relevant stock prices. As of January 2010, there were 66 financial companies to model, with the freshest readings being the close (adjusted for dividends and splits) prices taken on December 31, 2009. (All relevant share prices were retrieved from http://www.finance.yahoo.com.) Some of the modeled companies do present deterministic and robust share price models. As before, those S&P 500 companies which started after 2004 are not included. In addition, we have modeled Fannie Mae and Freddie Mac, which are not in the S&P 500 list, and Lehman Brothers and CIT Group (CIT) which are out of the S&P 500 list. Due to the fact that the latter three companies are both bankrupts, they have been modeled over the period of their existence. Apparently, there are many more bankrupts to be modeled in the future.

There are two sources of uncertainty associated with the difference between observed and predicted prices, as discussed by Kitov (2010). First, we have taken the monthly close prices (adjusted for splits and dividends) from a large number of recorded prices: monthly and daily open, close, high, and low prices, their combinations as well as averaged prices. Without loss of generality, one can randomly select for modeling purposes any of these prices for a given month. By chance, we have selected the closing price of the last working day for a given month. The larger is the fluctuation of a given stock price within and over the months the higher is the uncertainty associated with the monthly closing price as a representative of the stock price.

Second source of uncertainty is related to all kinds of measurement errors and intrinsic stochastic properties of the CPI. One should also bear in mind all uncertainties associated with the CPI definition based on a fixed basket of goods and services, which prices are tracked in few selected places. Such measurement errors are directly mapped into the model residual errors. Both uncertainties, as related to stocks and CPI, also fluctuate from month to month.

2. **Modeling financial companies**

The results of modeling are presented in Table 1 and Appendix 2: two defining components with coefficients and lags, linear trend and free terms, and the standard error, σ, expressed in dollars. Negative lags, which correspond to leading share prices, are shown in bold. Figure 1 and Appendix 3 depict the observed and predicted curves. Five companies will be studied in more detail in Section 5: American International Group, Citigroup (C), Fifth Third Bancorp (FITB), Legg Mason Inc. (LM), Moody's Corporation (MCO) and Morgan Stanley (MS).

Some financial companies have at least one defining CPI component lagging behind relevant stock price. For these companies, it is better to use the term "decomposition into" instead of "defining" CPI components. For example, share price of Aflac Incorporated (AFL) is defined by the index of financial services (*FS*) and that of transportation services (*TS*), the former lagging 2 months behind the share price and the latter leading by 6 months. Coefficient $b_1$ is positive. It means that the higher is the price for financial services the larger is the AFL's share price. The effect of the price index of transpiration services is opposite. Standard error for the model for the period between July 2003 and December 2009 is only $3.71. Figure 1 displays the observed and predicted prices for the period between 2003 and 2010. Before July 2003, the model does not hold and the curves deviate. Otherwise both curves are is a relatively good agreement including the sharp drop in 2008. From the statistical point of view, this is a key feature because any increase in the range of total change in the price and the defining CPIs is directly converted into higher model resolution.

Overall, standard errors in Table 1 and Appendix 2 vary from $0.77 for People's United Financial Inc. (PBCT) to ~$92 for AIG, which will be thoroughly analyzed in Section 5. When normalized to the stock prices averaged over the whole period, the standard errors fluctuate less. However, for non-stationary time series with measurement errors dependent on amplitude the normalized errors are likely biased. The predicted curve is Figure 1 is very close to the observed one and foresees one month ahead. Actually, the predicted curve leads the observed one by one month.

American International Group was the first company bailed out by the US financial authorities in September 2008. This action introduced a bias into the link between AIG share price and defining CPIs, which existed before September 2008. The model listed in Table 1 is likely to be inappropriate as related to the link and not a robust one. The defining CPIs for the December 2009 model are as follows: the index for food away from home (*SEFV*) leading by 1 month and the index of prescribed drugs (*PDRUG*) leading by 13 months. In Section 5, we investigate the evolution of the bet-fit model from May 2008 to December 2009.

The model for Apartment Investment and Management Company (AIV) has both defining CPIs leading the share price: the index of pets and pet related products (*PETS*), a subcategory of the index for recreation, leads by one month and the index of prescribed drugs (*PDRUG*) is five months ahead of the price. At first glance, this set of defining CPIs does not look convincing. This might be an effect of the changing trend in the CPI. Before November 2009, the best-fit model included the index of food and beverages (*F*) and the *PDRUG,* both leading by 8 months. This set determined the best-fit model during the 12 previous months (Kitov, 2010). In the smaller set of 34 CPI components used by Kitov (2010), the index of food and beverages and that of medical care (*M*) were the driving ones between November 2008 and October 2009 and provided the standard error of $2.058, but for a shorter period. With the *PDRUG*, the standard error for the same period between July 2003 and October 2009 is $2.055, i.e. only marginally better. This fact demonstrates how sensitive the model is to the defining CPIs. When more components are included, one could expect changes in the previously obtained models and lower standard errors.

The Allstate Corporation (ALL) has a model with both defining CPIs leading the share price. This model is unstable, however, and minimizes the RMS error only for the period between July 2003 and December 2009. In 2009, one of two defining components was randomly changing and one, the index for food away from home (SEFV), fixed. It is likely that the current set of defining CPIs do not include the one related to ALL. Thus, further investigations are needed.

Table 1. Empirical 2-C models for selected S&P 500 financial companies

| Company | $b_1$ | $CPI_1$ | $\tau_1$ | $b_2$ | $CPI_2$ | $\tau_2$ | $c$ | $d$ | $\sigma$, $ |
|---|---|---|---|---|---|---|---|---|---|
| AFL | 0.59 | FS | -2 | -2.37 | TS | 6 | 12.12 | 349.73 | 3.71 |
| AIG | -191.36 | SEFV | 1 | 38.53 | PDRUG | 13 | 727.81 | 21116.78 | 92.3 |
| AIV | -1.64 | PETS | 1 | 1.09 | PDRUG | 5 | -1.61 | -139.18 | 2.15 |
| ALL | 0.07 | E | 11 | -6.86 | SEFV | 2 | 45.20 | 1106.19 | 2.82 |
| AVB | 0.57 | CM | 1 | 1.92 | AB | -1 | -12.01 | -345.48 | 1.36 |
| AXP | -3.81 | F | 4 | -2.00 | M | 10 | 49.93 | 1115.79 | 2.48 |
| BAC | -2.95 | FB | 3 | -2.97 | SEFV | 13 | 35.43 | 956.52 | 2.53 |
| BBT | -1.57 | F | 3 | -0.31 | FRUI | 13 | 12.58 | 332.95 | 2.06 |
| BEN | -7.95 | FB | 3 | 6.59 | VAA | 13 | 60.58 | 564.48 | 6.46 |
| BK | -0.69 | MEAT | 13 | -1.65 | PETS | 1 | 15.80 | 270.32 | 2.09 |
| BXP | 4.58 | MCC | 5 | -5.04 | PETS | 3 | 5.63 | -605.31 | 5.04 |
| C | 2.54 | HO | 5 | -8.26 | SEFV | 2 | 36.70 | 1048.90 | 2.53 |
| FITB | -4.85 | SEFV | 2 | 1.45 | HS | 6 | 21.19 | 621.56 | 1.82 |
| HBAN | -2.15 | RPR | 13 | 1.32 | FOTO | 13 | 17.23 | 252.80 | 0.93 |
| HCN | -1.80 | PETS | 2 | 0.76 | HOSP | 5 | -7.06 | -40.40 | 2.20 |
| GS | 21.06 | HO | 10 | -29.45 | SEFV | 3 | 111.40 | 2496.20 | 13.48 |
| JPM | -2.49 | F | 4 | 3.19 | ORG | 0 | 26.31 | 139.00 | 2.49 |
| L | -2.49 | FB | 5 | -1.51 | TS | 3 | 28.35 | 679.85 | 2.07 |
| LM | -6.01 | F | 4 | -8.17 | APL | 13 | 33.07 | 1754.81 | 6.89 |
| PNC | 1.49 | CM | 0 | -3.44 | FB | 4 | 16.37 | 331.72 | 3.49 |
| PSA | -4.14 | SEFV | 3 | 2.04 | PDRUG | 5 | 14.25 | 72.98 | 4.43 |
| VNO | -11.08 | SEFV | 3 | 2.23 | PDRUG | 5 | 57.23 | 1113.14 | 5.30 |

Avalonbay Communities (AVB) has a model with one defining index (alcoholic beverages, *AB*) lagging behind the price by one month and the headline CPI less medical care (*CM*) leading by one month. This model is very stable over the previous 10 months and has a standard error of $1.36.

American Express Company (AXP) has a model predicting at a four month horizon. The defining CPIs are the index for food and beverages leading by 4 months and the index for medical care leading by 10 months. In the previous study (Kitov, 2010) the model was essentially the same. So, the extended CPI set does not make a better model. The model is a robust one and minimizes the standard error for the period between July and November 2009 as well.

The model for Bank of America (BAC) is defined by the index of food and that of food away from home. The latter CPI component leads by 13 months. From our past experience, the larger is the lag the more unreliable is the model. However, both defining components provide the best fit model in the second half of 2009. Both coefficients in the BAC model are negative. This means that increasing food price forces the share price down. The growth in the indices of food and food away from home has been compensated by linear time trend in the share price.

Franklin Resources (BEN) is driven by the index of food (*FB*) and that of video and audio (*VAA*), both leading by several months. The former component has a negative coefficient and the latter one – positive. Bank of New York Mellon Corporation (BK) is defined by the index of meats, poultry and fist (*MEAT*) and the index of pets and per related products, both having negative coefficients. The model has a standard error of $2.09.

Boston Properties (BXP) has a model with the index of medical care commodities (*MCC*) and *PETS* leading by 5 and 3 months, respectively. This is a relatively stable model. However, the best-fit model was different before September 2009 and included the index of food and the index of miscellaneous services (*MISS*). This model had been reining since March 2009. The model obtained in (Kitov, 2010) was based on the CPI less energy (*CE*) and the index of food. It was a mediocre model with the RMS error of $5.54 compared to $5.11 obtained in this study for the same period.

Citigroup is of special interest. This company was bailed out in November 2008. For the purposes of share modeling the bailout introduces a major disturbance, because the share is not the one estimated by the free stock market any more. Accordingly, the models obtained after November 2008 are likely to be biased. In Table 1, a share of Citigroup is defined by the index of household operations (HO) and that of food away from home (SEFV). Coefficient $b_1$ (=+2.54) is positive and the increase in *HO* should be converted into a higher share price. The effect of *SEFV* is an opposite one with a larger coefficient $b_2$=-8.26. In 2007 and 2008, the index of household operations was increasing at almost the same rate as the *SEFV*, and the share fell from $50 in April 2007 to $1.5 in January 2009.

The next ten companies in Table 1 are all robust and have deterministic price models with no one of defining indexes lagging. Huntington Bancshares (HBAN) is controlled by the index of primary residence rent (*RPR*) and the index for photography (*FOTO*), both leading by 13 months. This model holds at least over the 10 months previous to December 2009. Coefficient $b_1$ is negative and any increase in *RPR* is converted into a decrease in the HBAN share price 13 months later. It is instructive to track this model in 2010. It must be fading away with the transition to a new trend in the CPI. The model standard deviation is only $0.92 for the whole period. Figure 1 displays relevant curves. Except four of five points, the agreement is excellent.

Health Care REIT (HCN) has a model defined by the PETS and the index of hospital services (*HOSP*). The latter has a permanent positive trend and likely is compensated by the linear trend term with a negative slope (-40.4). The predicted and observed curves are very close. So, the model accurately predicts at a two-month horizon.

Goldman Sachs (GS) is a famous bank. The trajectory of its share price is well predicted by the index of household operations (*HO*) and the index of food away from home (*SEFV*) at a three month horizon. Since coefficient $b_1$ is positive the decreasing price index of household operations results in a fall in GS share price, as was observed in 2008 and 2009. In the second half of 2009, the price was on rise.

JPMorgan Chase & Co. is defined by food and beverages (*F*) and other recreation goods (*ORG*). Since the time lag of the *ORG* is zero the model can predict only contemporary share price. In Figure 1, the observed and predicted curves almost coincide before 2007. The years between 2007 and 2009 are characterized by extremely high fluctuations in the observed price. The model failed to predict this feature. In 2009, the prediction is good again, however. Therefore, the fluctuations are likely to be related to short-term forces not affecting fundamental long-term link between the price and the defining CPIs.

The best-fit model for Loews Corporation (L) includes the index of food (*FB*) leading by 5 months and the index of transportation services (*TS*) leading by 3 months. Both coefficients are negative and are counteracted by a positive slope *c*=28.35. The model for Legg Mason (LM) is based on the index of food and beverages and the index of appliances (*APL*) from the housing index, the latter leading by 13 months. Overall, the predicted time series is very close to the observed one with standard deviation of $6.89. The largest input to the standard deviation comes from a short period in 2006. Otherwise, both curves are very close even during the dramatic fall from $80 per share in the end of 2007 to $10 per share in February 2009 and during the fast recovery in 2009.

PNC Financial Services (PNC) relies on the headline CPI less medical care (*CM*) and the index of food (*FB*), the model being a contemporary to the share. Public Storage (PSA) and Vornado Realty Trust have similar models defined by the index of food away from home and the *PDRUG*. The time lags are also identical and are 3 and 5 months, respectively. Figure 1 demonstrates that the observed prices of PCA and VNO are similar with a local peak in the first half of 2008. A similar pattern is observed for AIV, which model also includes *PDRUG*. The difference between PCA and VNO is in the sensitivity to *SEFV*: $b_1$(PSA)=-4.14 and $b_1$(VNO)=-11.08.

So, among the models with both defining CPIs leading relevant shares, there are examples of robust models and unstable models. For the latter companies, no fixed model is available over the past year. It is likely that these models express the lack of true defining indices in the current set of CPIs and are affected by random measurement noise. One cannot exclude that true robust models do exists for these companies.

For other forty four financial companies relevant models and graphs are presented in Appendices 2 and 3. These models use quiet several defining CPIs not mentioned in Table 1. Otherwise, Table 1 contains all meaningful configurations of leading and lagging share prices and those in the Appendices are given for the sake of completeness.

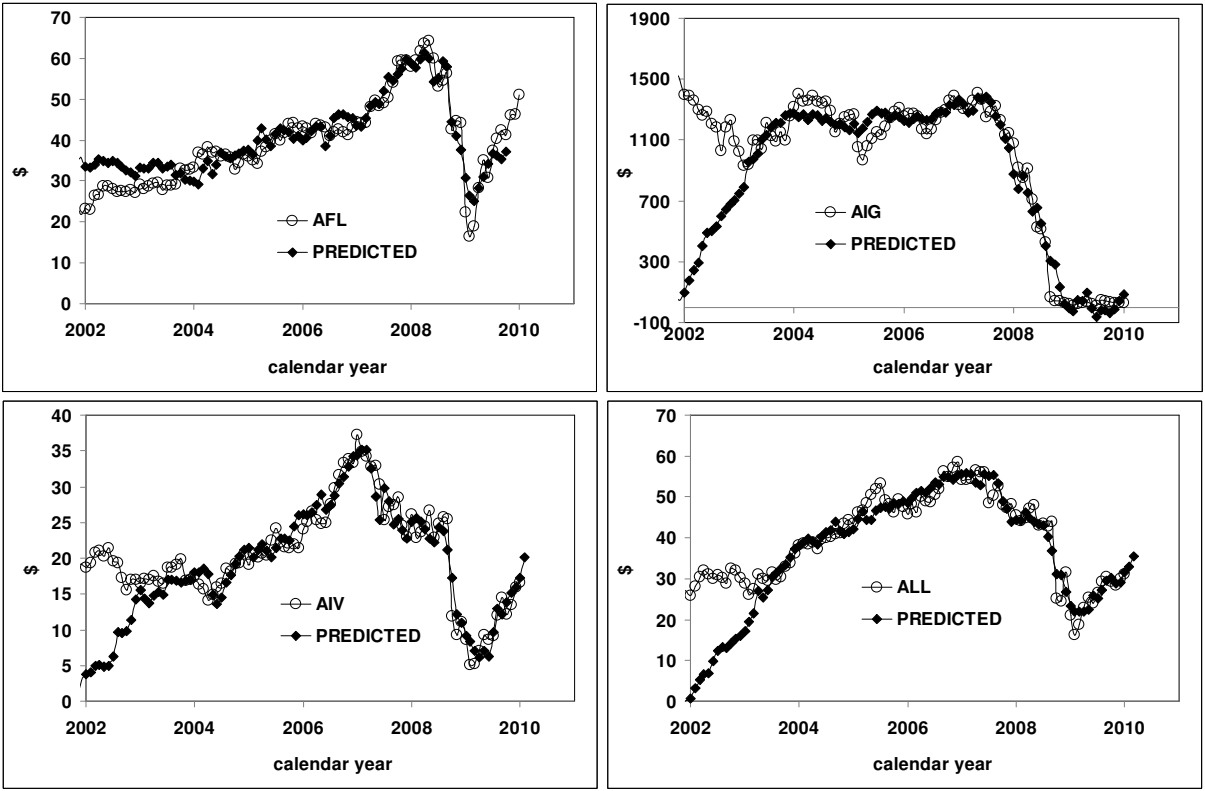

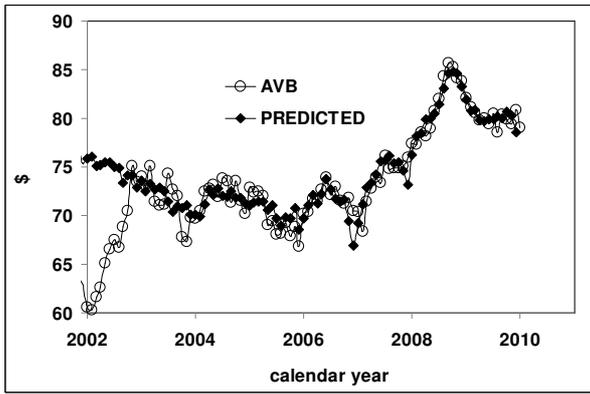
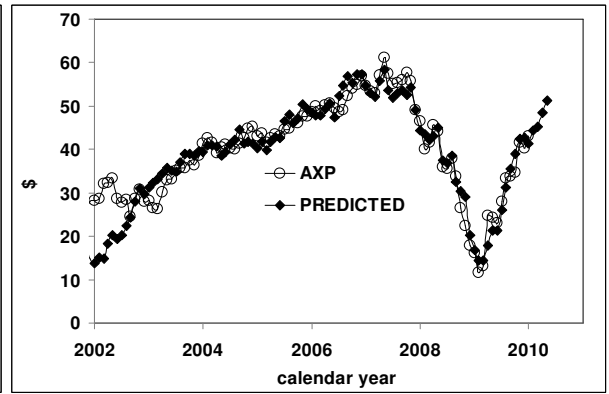
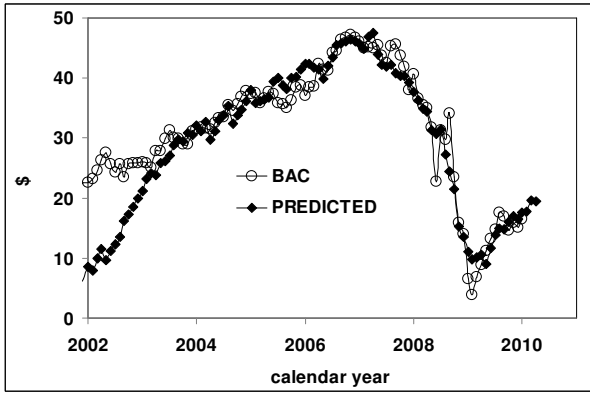
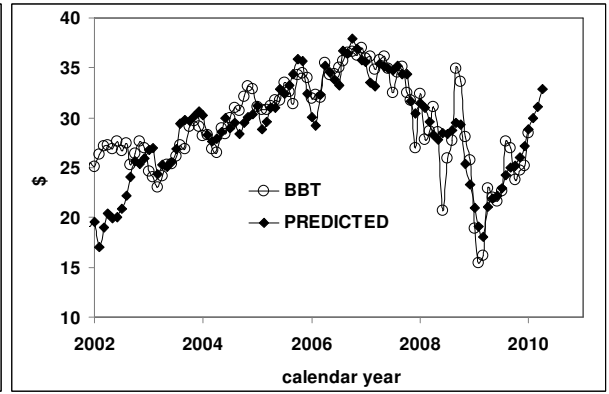
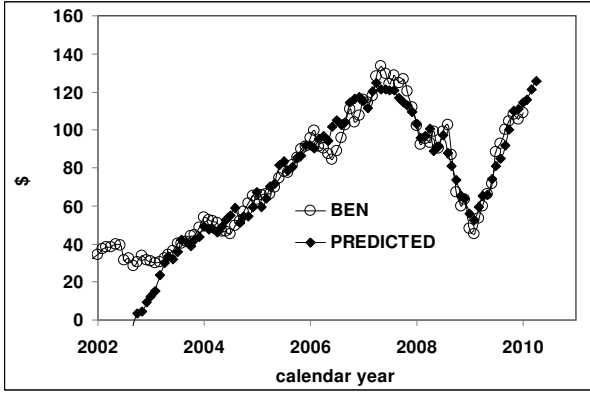
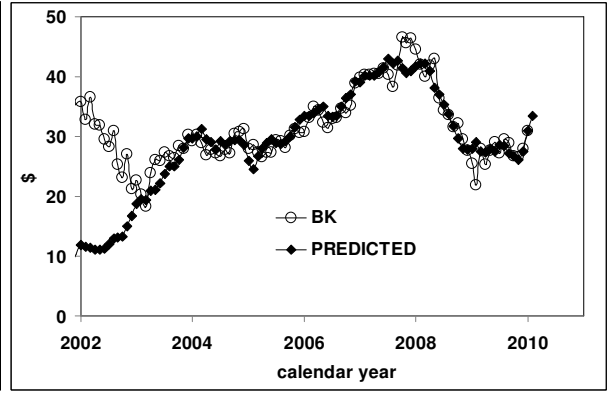
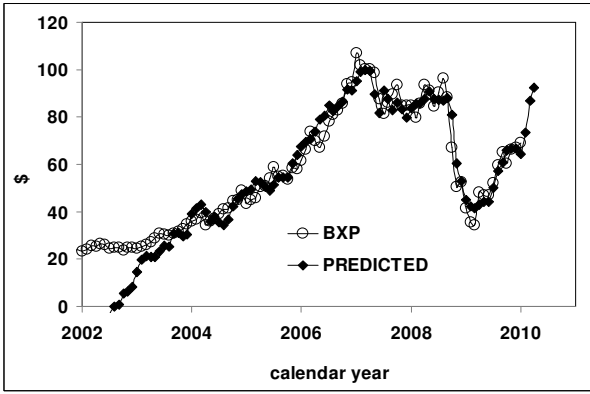
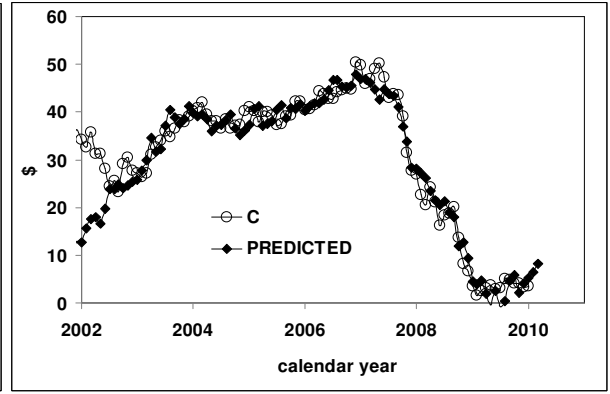

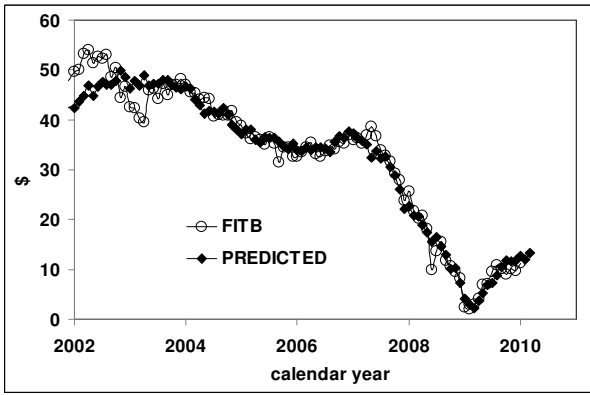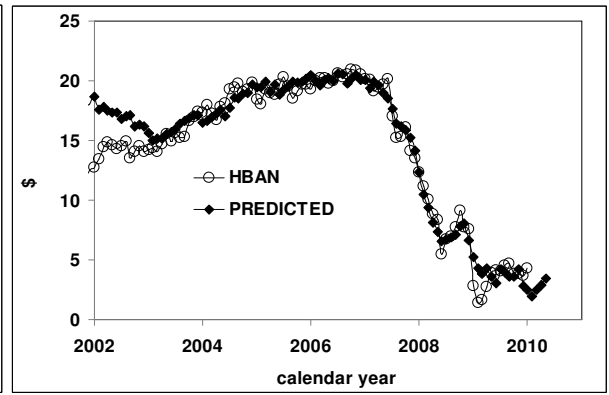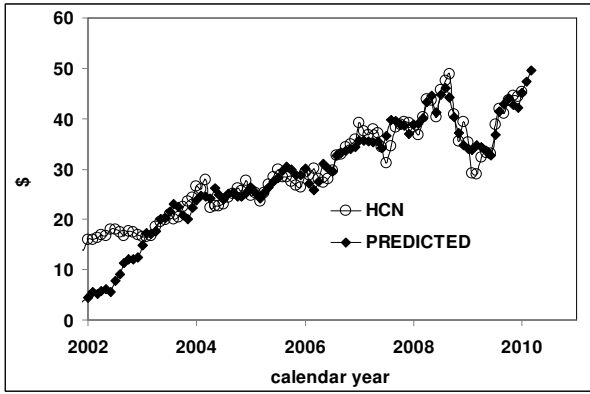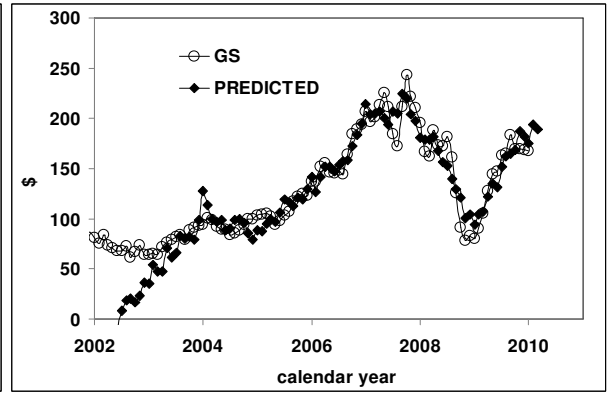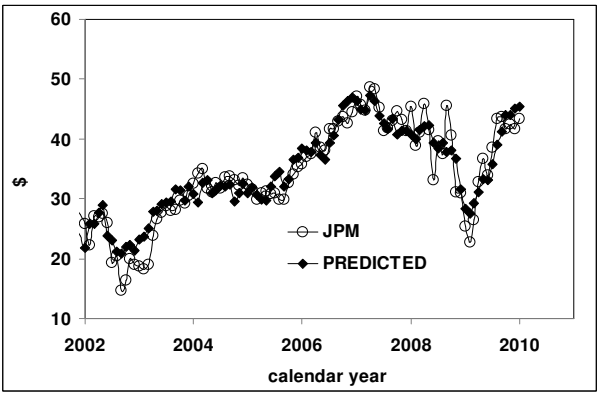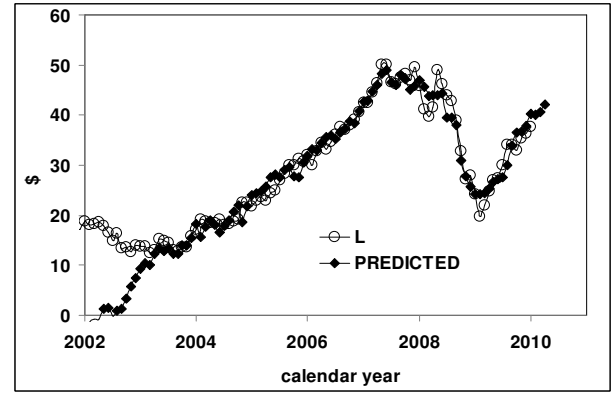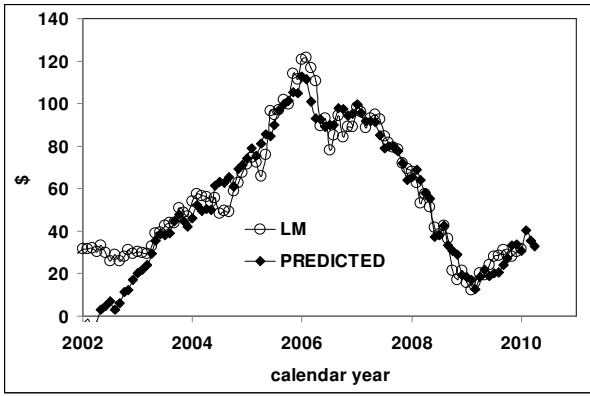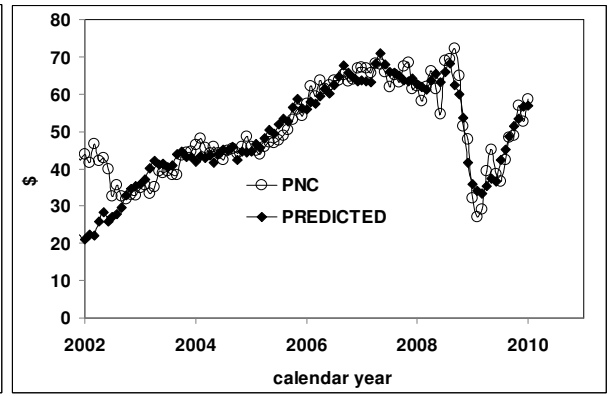

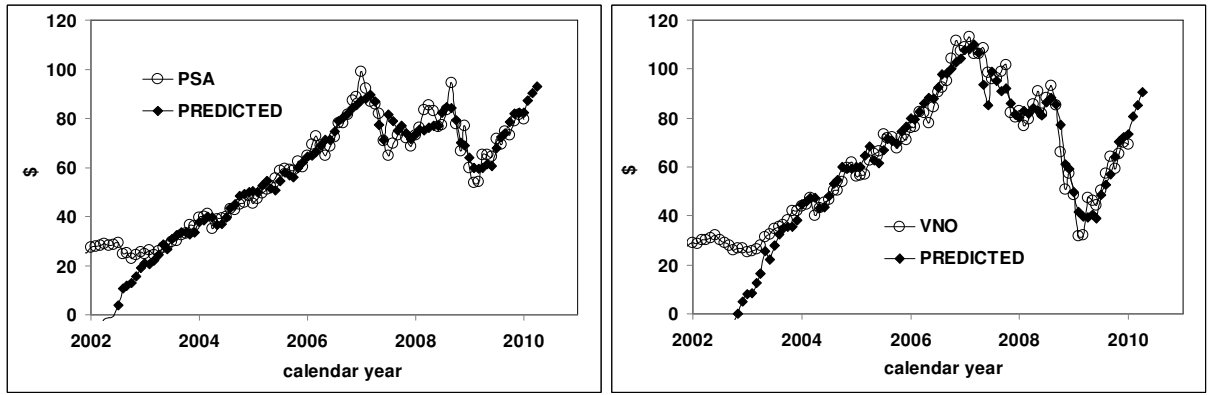

Figure 1. Observed and predicted share prices of eight financial companies from the S&P 500 list. Relevant empirical models are presented in Table 1.

The principal purpose of Section 2 consists in presenting tentative empirical models for share prices of financial companies. The current set of defining CPIs is far from a complete one and further investigation may reveal more accurate and reliable models for the same companies. However, the current models might be good enough because of high correlation between various CPI components. For a given share, the currently used CPIs may be very close to the true defining CPIs, which are not included in the set yet. Therefore, a direct statistical estimate of the model accuracy and reliability is a major task.

## 3. Cointegration tests

Statistical properties of the residual error are crucial for any quantitative model. Ideally, a good model involving time dependent measured variables should describe the essence of real ties. The model residual error should be a stationary (and thus, integrated of order zero) stochastic process with characteristics close to the uncorrelated white noise with Gaussian distribution. In the best case, residual errors should depend only on measurement errors, with the measurements conducted in a consistent way.

As in the previous study (Kitov, 2010), we applied the Johansen cointegration test to the observed time series and those predicted in Section 2. For all studied companies, the test resulted in cointegration rank 1 or, for two non-stationary variables, in the presence of one cointegrating relation. All results are listed in Table 2. The Johansen approach does not require both variables to be in the same order of integration.

As an alternative, we have applied the augmented Dickey-Fuller (ADF) and the Phillips-Perron (PP) tests for unit roots to all residual errors of the models in Table 1, except those with defining CPI lagging behind relevant shares. This procedure is in line with the Granger-Engle two-step method based on several specification tests applied to the residual time series. Having the same econometric power as the Johansen procedure, the Granger-Engle test allows a larger variety of specifications.

Table 2. Results of Johansen cointegration test and unit root tests as applied to the residual errors

|  | ADF  z(t), 1%CV[1]=-3.54 | PP  z(ρ), 1%CV=-19.4 | Johansen test | | |
|---|---|---|---|---|---|
|  |  |  | eigenvalue | trace statistics, 5%CV=3.76 | rank |
| AIG | -4.10 | -29.8 | 0.21 | 0.24 | 1 |
| AIV | -6.19 | -47.5 | 0.35 | 0.15 | 1 |
| ALL | -5.95 | -46.5 | 0.32 | 0.04 | 1 |
| AXP | -4.98 | -38.3 | 0.25 | 0.35 | 1 |
| BAC | -4.87 | -38.5 | 0.21 | 0.003 | 1 |
| BBT | -5.96 | -47.7 | 0.29 | 0.08 | 1 |
| BEN | -4.53 | -34.0 | 0.22 | 1.90 | 1 |
| BK | -5.00 | -40.0 | 0.26 | 0.07 | 1 |
| BXP | -5.47 | -39.6 | 0.33 | 0.87 | 1 |
| C | -5.60 | -41.0 | 0.36 | 0.27 | 1 |
| FITB | -5.63 | -45.8 | 0.26 | 2.39 | 1 |
| GS | -4.56 | -34.0 | 0.41 | 1.69 | 1 |
| HBAN | -5.44 | -41.5 | 0.29 | 0.29 | 1 |
| HCN | -5.27 | -39.0 | 0.37 | 1.91 | 1 |
| JPM | -6.22 | -51.2 | 0.30 | 1.15 | 1 |
| L | -5.70 | -43.5 | 0.37 | 0.96 | 1 |
| LM | -4.75 | -37.2 | 0.18 | 0.11 | 1 |
| MS | -6.20 | -48.4 | 0.33 | 0.09 | 1 |
| PNC | -6.12 | -50.0 | 0.33 | 0.47 | 1 |
| PSA | -5.84 | -44.3 | 0.33 | 2.18 | 1 |
| VNO | -5.46 | -40.0 | 0.33 | 0.83 | 1 |

[1]CV – critical value

In a sense, this Section is a fully technical one. We need only a confirmation that the regression technique used in Section 2 is applicable, i.e. the regression does not give spurious results. Both tests for cointegration unambiguously evidence the presence of long-term equilibrium relations between the actual and predicted prices. The predicted prices can be considered as weighted sums of prices for goods and services. In this regard, they are similar to the overall CPI and can be considered as independent measurements and represent just one variable. Therefore, one does not need to test both defining CPI for cointegration with relevant share price.

So, one can derive a conclusion that the deterministic pricing model provides a statistically and econometrically valid description of share prices of S&P 500 financial companies. There is a problem with the model resolution, however. As happens often in physics, in order to obtain a consistent and reliable model one should have a wider dynamic range of involved variables or to increase the accuracy of measurements. The latter is hardly possible with the past CPI readings. So, one could expect a more reliable model for the companies with share prices varying the most.

## 4. May 2008 vs. December 2009

The current models predicting future prices are of crucial interest for the stock market. It is always important to know which stocks will go up/down and at what rate. However, there are significant problems related to the past performance of the stock market also to be considered. One of these problems is associated with the 2008/2009 financial and economic crisis, which exposed many companies to major risks. Since the late 2007 and very actively since July 2008, stock prices of many companies have been decreasing at an accelerating speed. The decrease costs trillions US dollars net lost after the overall asset devaluation. This is a natural challenge to our concept: Could the model predict the fall in stock prices if available in 2008?

For all investors and owners it would have been a great relief to predict, and thus, prevent or reduce the loss. Here we would like to stress again that the model is valid only when it does not disturb natural functioning of the stock market, i.e. those myriads of well-established direct and indirect interactions between economic and financial agents. When everybody shifts to one or few "salvage" stocks, their behavior becomes highly distorted, biased, and thus unpredictable. A part of the financial market is never equivalent to the whole market and this model will be worthless when used by all market players. So, we would not recommend using the model shortly after this book is published. In a sense, this publication may destroy the market configuration described by the model.

The principal question posed in this Section can be addressed quantitatively. As a first step, we move back in May 2008 and use contemporary CPI data to obtain the best-fit models for the S&P 500 share prices under study. Table 3 and Appendix 4 list these models obtained for selected financial companies. One should bear in mind that the involved prices had only limited dynamic range in the beginning of 2008 and corresponding models are not fully resolved. In this sense, the 2009 models are superior.

Then, we calculate all share prices using the 2008 models and actual CPI data between May 2008 and December 2009. In Figure 2 we compare the 2008 predictions to those obtained in January 2010 (i.e. the models for December 2009) and described in Section 2. If both models for a given share provide similar predictions then the 2008 fall was predictable, at least for the company.

A few companies in Table 3 have one defining component of the same nature as that in relevant December 2009 models. For a majority, both defining indices are different. This effect is observed despite our intention to select those 2008 models which provide the best prediction.

The May 2008 model for ALL, which includes the index of diary products (*DIAR*) and the index of intracity transportation (*ITR*), predicts the evolution of the share price relatively well till July 2009. Then the May prediction starts to diverge at an accelerating rate from the observed trajectory. The 2008 trough could be forecasted both in time and amplitude in May 2008.

Table 3. Defining CPI components, coefficients and time lags for the models in May 2008.

| Company | $b_1$ | $CPI_1$ | $\tau_1$ | $b_2$ | $CPI_2$ | $\tau_2$ | c | d |
|---|---|---|---|---|---|---|---|---|
| AFL | 0.36 | DIAR | 0 | -0.55 | ITR | 12 | 8.92 | 38.50 |
| AIG | -21.11 | DIAR | 9 | -172.66 | **SEFV**[1] | 2 | 1148.11 | 31872.06 |
| AIV | 1.78 | VAA | 3 | 1.23 | **PDRUG** | 7 | -10.77 | -525.90 |
| ALL | 1.85 | MCC | 7 | -2.22 | APL | 9 | -10.18 | -207.35 |
| AVB | 1.41 | EC | 4 | 1.59 | **AB** | 1 | -10.47 | -340.92 |
| AXP | -7.87 | SEFV | 4 | 1.20 | HS | 6 | 47.73 | 1098.50 |
| BAC | -2.72 | **F** | 3 | 1.77 | TS | 12 | 8.86 | 115.22 |
| BBT | -1.98 | RPR | 13 | -1.16 | MISS | 12 | 25.57 | 648.22 |
| BEN | 3.28 | PDRUG | 12 | -1.90 | HOSP | 3 | 30.13 | -368.96 |
| BK | -1.44 | H | 11 | 3.13 | HO | 11 | -2.39 | -75.70 |
| BXP | -3.96 | F | 4 | 3.17 | **MCC** | 7 | 14.55 | -136.73 |
| C | -4.02 | FB | 8 | -4.00 | RPR | 13 | 46.97 | 1376.82 |
| FITB | -0.40 | DIAR | 7 | 0.84 | PDRUG | 8 | -12.03 | -112.79 |
| GS | 12.87 | MVP | 5 | 27.41 | FOTO | 12 | 65.74 | -4221.10 |
| HBAN | -2.07 | SEFV | 6 | -1.66 | **RPR** | 13 | 22.63 | 638.78 |
| HCN | 0.20 | FU | 8 | -2.27 | VAA | 0 | 1.07 | 225.28 |
| HST | -2.81 | SEFV | 3 | 0.55 | PDRUG | 7 | 13.60 | 290.45 |
| JPM | -2.54 | **F** | 4 | -0.45 | MEAT | 13 | 20.29 | 484.70 |
| L | 2.02 | MVP | 5 | -5.37 | SEFV | 5 | 33.34 | 643.83 |
| LM | -2.56 | MEAT | 7 | -12.00 | **APL** | 13 | 12.38 | 1528.90 |
| PNC | 2.43 | DUR | 10 | -5.69 | SEFV | 4 | 43.97 | 617.15 |
| PSA | 6.13 | MVP | 5 | -9.54 | RPR | 6 | 57.57 | 1103.99 |
| VNO | 2.63 | **PDRUG** | 7 | 8.56 | FOTO | 10 | 10.67 | -1683.89 |

[1] same defining CPIs as in Table 1 are highlighted.

For several companies, the agreement between the May 2008 and December 2009 predictions does not disappear even when models are different. For example, the 2008 model for AXP, defined by the *SEFV* and the index of household services (*HS*), does not diverge from the observed trajectory since 2008. Similar situation is observed with the model for Host Hotels&Resorts (HST). This effect shows the necessity of a complete or at least more representative set of CPIs. Otherwise, one can not distinguish between two neighboring models with defining CPI components characterized by a high degree of correlation.

The 2008 and 2010 models for BXP have one common defining variable – the index of medical care commodities (*MCC*). Nevertheless, the 2008 model fails to predict the future trajectory well. This is due to the difference between the indices for food and beverages and pets. Similar effect is observed with JPM.

The 2008 model for GS demonstrates a striking difference with the observed time series. It predicted the fall in the share price down to the zero line in the second half of 2009. In reality, no catastrophic drop happened and the price fell only to the level of $70. From the actual time series it is clear that the model for GS could not be well resolved because of very limited change in the share

price by May 2008. There are several financial companies with shares predicted to fall below the zero. Some predictions were accurate enough. These companies are modeled in Section 5.

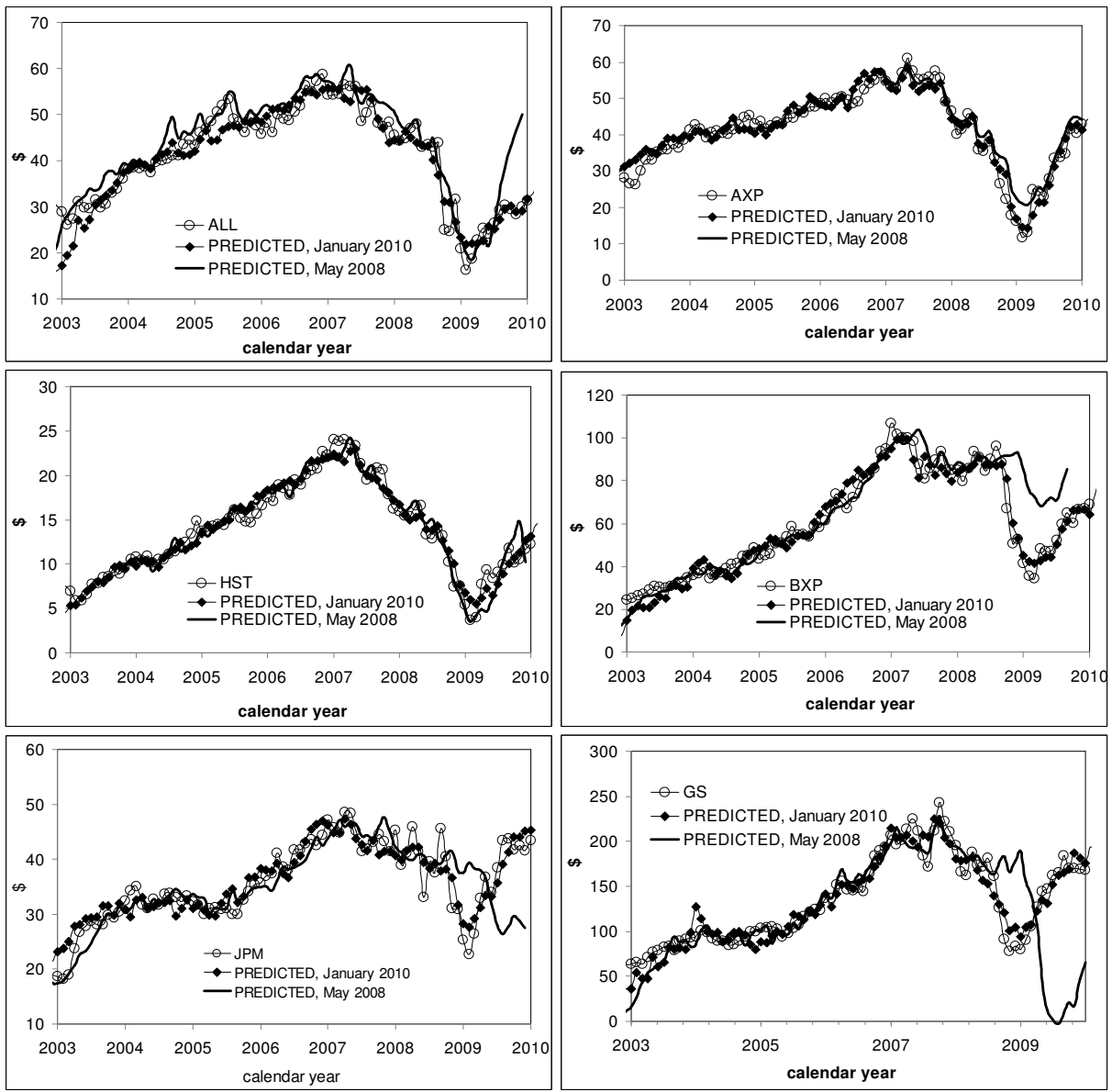

Figure 2. Comparison of stock prices predicted in May 2008 and December 2009.

The number of successful models is relatively small if to consider the initial set of S&P 500 companies. This fact raises delicate questions about the reliability of the models and the concept itself. One may assume that the successful models are just a lucky accident. The concept should be validated by modeling of more companies, extension of the set of defining CPI components, and usage of new data in the years to come.

We also take into account the fact that quantitative models, also in physics, are better resolved than all involved variables vary in wider ranges. Specifically, the difference between the 2008 and 2009 models consists in the sharp fall after July 2008. Therefore, the models obtained in 2009 are better resolved and thus superior to those from 2008. Data available in 2008 did not allow identification of right models because of high correlation between subcategories of the consumer price index. Good news is that the right models hold once and for all, but with new coefficients.

## 5. Predicting bankruptcy

In Section 4, we have modeled the evolution of share prices of several financial companies from the S&P 500 list between May 2008 and December 2009. It was found that some predicted share prices sank below the zero line. Under our framework, the presence of a negative stock price may be considered as an equivalent to a net debt. When long enough and without any positive prospective, such a debt would likely result in a bankruptcy.

In reality, some companies with negative predicted share prices declared bankruptcy, some were bailed out and some have been suffering tremendous difficulties since 2008. The first group is represented by Lehman Brothers who filed for Chapter 11 bankruptcy protection on September 15, 2008. The net bank debt was estimated at the level of $600 billion. More than 100 banks filed for bankruptcy since then.

Several banks were bailed out, with American International Group the first to obtain a $150 billion government bailout. The AIG bailout was presented as a major move to save the collapsing US financial system. The biggest examples of bailout are also Fannie May and Freddie Mac. All three companies had a sharp share price fall in the second half of 2008.

CIT Group Inc. (CIT) got $2.3 billion of bailout money in December 2008 and $3 billion bond holder bailout in July 2009. However, it did not help and CIT declared bankruptcy in November 2009. These companies and many others have been struggling and likely will struggle in the future trying to restructure their debts and re-enter the stock market.

Section 5 seeks to answer a number of questions:
- Was it possible to predict the evolution of total debt of the bankrupts?
- Was it possible to predict the dates of these bankruptcies?
- Is it possible to predict the date of recovery?
- It is possible to predict future bankruptcies?
- Which company had to be bailed out and when?

All S&P 500 models with negative share prices were obtained together with other models for May 2008. In this regard we should not distinguish them. The reason for a separate investigation consists in the fact that negative share prices might result in bankruptcies. This is a phenomenon no described quantitatively by our models and thus deserving special attention. Otherwise, all models were equivalent and obtained according to the same procedures. It is worth noting that the models for the same companies obtained in October 2009 are highly biased by bailouts or do not exist together with bankrupt companies.

Table 4. Models for 10 companies: May, September and December 2008, and October 2009.

May 2008

| Company | $b_1$ | $CPI_1$ | $\tau_1$ | $b_2$ | $CPI_2$ | $\tau_2$ | $c$ | $d$ |
|---|---|---|---|---|---|---|---|---|
| AIG  | -21.11 | DIAR | 9 | -172.66 | SEFV  | 2  | 1148.11 | 31872 |
| C    | -4.33  | FB   | 4 | -3.63   | RPR   | 12 | 46.79   | 1358  |
| CIT  | -4.84  | F    | 5 | 11.51   | SEFV  | 6  | 96.99   | 2610  |
| FITB | 1.46   | MCC  | 9 | -0.32   | DIAR  | 8  | 13.01   | 227.5 |
| FNM  | 9.62   | RS   | 3 | 10.84   | SEFV  | 6  | 24.36   | 733.0 |
| FRE  | -3.54  | DUR  | 2 | -9.66   | RPR   | 13 | 57.75   | 2180  |
| LEH  | -6.27  | FB   | 4 | -1.38   | HOSP  | 3  | 77.60   | 1411  |
| LM   | -2.57  | MEAT | 7 | -12.02  | APL   | 13 | 12.40   | 1532  |
| MCO  | -5.50  | F    | 5 | -5.83   | RPR   | 9  | 75.37   | 1909  |
| MS   | 7.788  | R    | 7 | -0.85   | DIAR  | 4  | 1.49    | -658.8 |

September 2008

| Company | $b_1$ | $CPI_1$ | $\tau_1$ | $b_2$ | $CPI_2$ | $\tau_2$ | $c$ | $d$ |
|---|---|---|---|---|---|---|---|---|
| AIG  | -22.10 | DIAR | 9  | -178.42 | SEFV  | 1  | 1198  | 32967  |
| C    | -4.26  | FB   | 9  | -3.62   | RPR   | 12 | 46.39 | 1345   |
| CIT  | -0.77  | DAIR | 8  | -8.20   | RPR   | 11 | 59.95 | 1584   |
| FITB | -3.07  | F    | 12 | 1.06    | PDRUG | 8  | -0.97 | 250.50 |
| FNM  | -15.39 | SEFV | 10 | 4.64    | HS    | 6  | 68.38 | 1937.7 |
| FRE  | -1.14  | COMM | 0  | -14.11  | SEFV  | 5  | 90.59 | 2433   |
| LEH  | -7.37  | FB   | 4  | -5.29   | MISS  | 2  | 102.3 | 2477.7 |
| LM   | -2.62  | MEAT | 7  | -12.20  | APL   | 13 | 12.46 | 1558   |
| MCO  | 3.28   | DUR  | 9  | -9.26   | RPR   | 9  | 72.78 | 1237.5 |
| MS   | -0.42  | TPU  | 0  | -0.95   | DIAR  | 4  | 12.95 | 235.2  |

December 2008

| Company | $b_1$ | $CPI_1$ | $\tau_1$ | $b_2$ | $CPI_2$ | $\tau_2$ | $c$ | $d$ |
|---|---|---|---|---|---|---|---|---|
| AIG  | -22.34 | DIAR | 9  | -173.7 | SEFV | 1  | 1169  | 32260  |
| C    | -3.74  | FB   | 9  | -3.76  | RPR  | 13 | 44.39 | 1287.6 |
| CIT  | 2.60   | NC   | 12 | -9.66  | RPR  | 13 | 63.46 | 1375   |
| FITB | -3.48  | F    | 7  | -0.93  | LS   | 11 | 23.20 | 781.5  |
| FNM  | -5.67  | F    | 8  | -2.28  | TS   | 0  | 35.61 | 1436   |
| FRE  | -2.21  | TS   | 0  | -8.40  | RPR  | 13 | 62.87 | 1976.6 |
| LEH  | -5.21  | F    | 5  | -4.97  | PETS | 0  | 59.65 | 1323   |
| LM   | -7.27  | F    | 5  | -8.31  | APL  | 13 | 39.48 | 1967.8 |
| MCO  | 2.98   | DUR  | 9  | -9.70  | RPR  | 10 | 74.58 | 1350   |
| MS   | -12.55 | SEFV | 3  | 2.83   | HS   | 10 | 74.88 | 1589   |

October 2009

| Company | $b_1$ | $CPI_1$ | $\tau_1$ | $b_2$ | $CPI_2$ | $\tau_2$ | $c$ | $d$ |
|---------|-------|---------|----------|-------|---------|----------|-----|-----|
| AIG  | -22.79 | DIAR | 9  | -156.05 | SEFV | 0  | 1066  | 29580 |
| C    | -0.59  | DIAR | 4  | -5.88   | SEFV | 5  | 38.84 | 1054.8 |
| CIT  | 4.92   | HFO  | 10 | -9.37   | RPR  | 12 | 62.02 | 1058 |
| FITB | -4.99  | SEFV | 2  | 1.54    | HS   | 6  | 21.86 | 630.7 |
| FNM  | -15.39 | SEFV | 10 | 4.64    | HS   | 6  | 68.38 | 1837.7 |
| FRE  | -1.13  | COMM | 0  | -14.12  | SEFV | 5  | 90.59 | 2433.8 |
| LEH  | -7.39  | FB   | 4  | -5.29   | MISS | 2  | 102.3 | 2477.7 |
| LM   | -5.82  | FB   | 4  | -8.18   | APL  | 13 | 32.36 | 1722 |
| MCO  | -12.98 | RPR  | 10 | 3.19    | MISG | 8  | 97.83 | 1981 |
| MS   | 5.16   | HO   | 10 | -9.61   | SEFV | 3  | 39.62 | 1017 |

Table 4 lists 10 models with predicted negative or very close to negative prices as obtained in May, September and December 2008 as well as in October 2009. Figure 3 displays corresponding predicted and observed curves between July 2003 and December 2009. American International Group has a very stable model for the entire period as defined by the *DIAR* and *SEFV*. Theoretically, the company should suffer a rapid drop in share price from ~$1400 to the level of about -$300. In reality, this fall was stopped by a bailout with the share price hovering between $10 and $50 by the end of 2008 and through 2009. According to all four models the price should start growing in 2010. It will be an important test for our pricing concept.

For Citigroup, the models obtained in 2008 are similar and are based on the indices of food and rent of primary residence. Figure 3 demonstrate that negative prices were expected in the end of 2008. All three models predicted the bottom price at -$30. In October 2009, the defining CPI components are different as the model tries to describe the price near $2.

The history of CIT Group (CIT) includes two attempts of bailout and a bankruptcy in November 2009 with a total debt of $10 billion. In Figure 3, the May 2008 model predicts a very deep fall in the share price. Other two models in 2008 demonstrate just a modest fall below the zero line. The bailouts have likely biased the October 2009 model and it predicts the company to recover in 2010. It would be a good exercise similar to that for the AIG model. Unfortunately, the history of CIT Group has ended with a bankruptcy, as expected.

Fanny Mae and Freddie Mac were both bailed out in September 2008. As depicts Figure 3, the models between May and December 2008 are all different. However, all of them predicted negative prices. The models for FNM imply the bottom price level of -$50 to -$60 and the pivot point somewhere in 2009. The models for FRE do predict negative prices with the bottom at -$30, but only the September model has a pivot point.

Lehman Brothers was one of the first giant companies to file for bankruptcy protection in September 2008. The May 2009 model does predict negative prices in the beginning of 2009. The September and December 2009 models are likely biased by the bankruptcy but both indicate a deep fall in the price. It is important to stress that the bottom price for LEH was predicted at -$20 with a quick return into the positive zone. Therefore, the risk might be overestimated.

The models predicted for FITB, LM, MCO and MS are presented to emphasize the problem of resolution and selection of a valid model. For these four companies there is at least one model predicting negative or very close to zero prices. In reality, no one of them has touched the zero line. Moreover, they have not been falling since the end of 2008. So, in order to obtain an accurate prediction one should the best resolution, which might be guaranteed by the higher possible dynamic range. The 2008 crisis and the following recovery allowed the biggest change in the S&P share prices. Hence, the models obtained in 2010 have to be the most resolved and thus the most reliable. Good news is that these models will be valid in the future, but with different coefficients (Kitov, 2010).

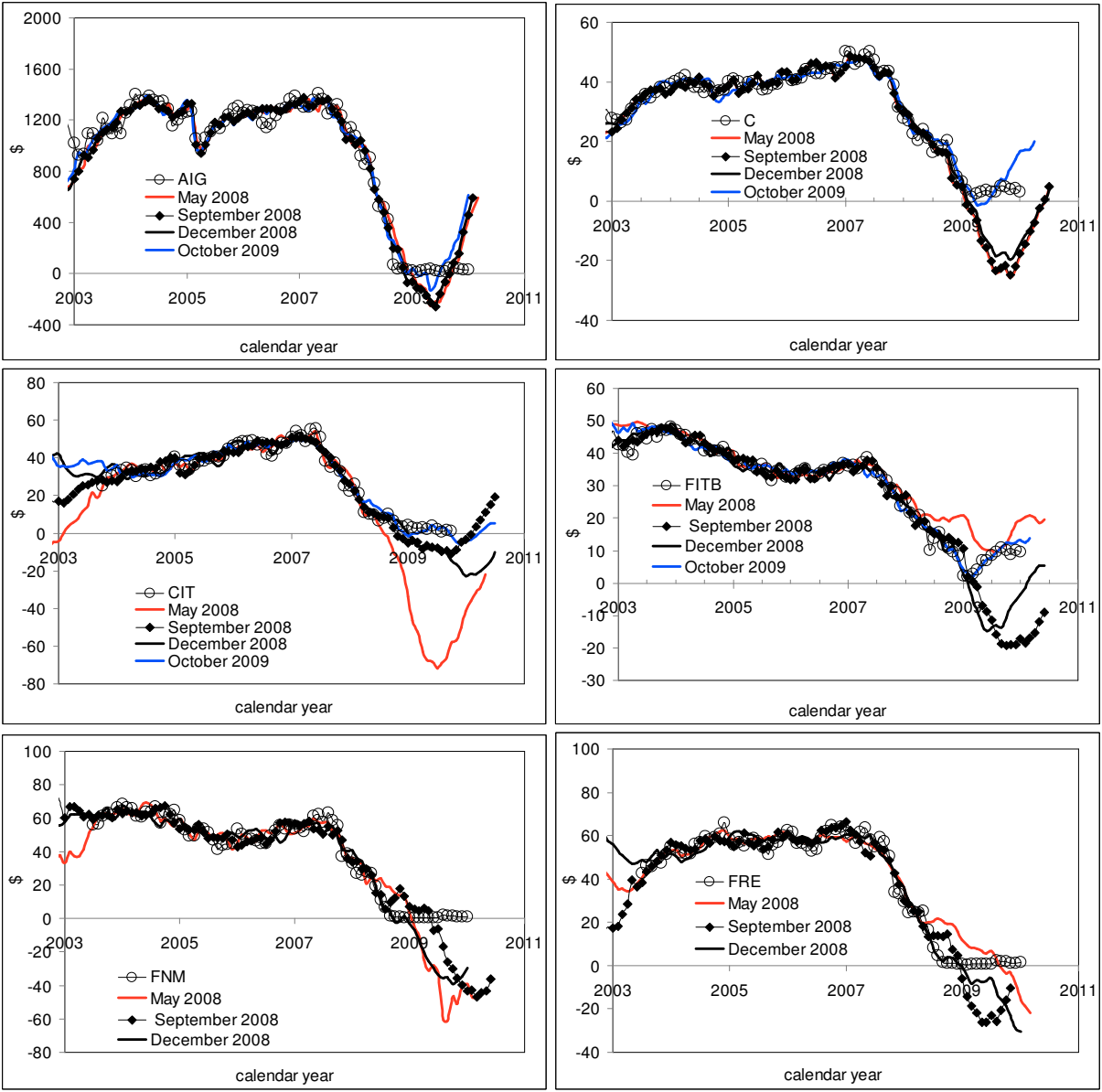

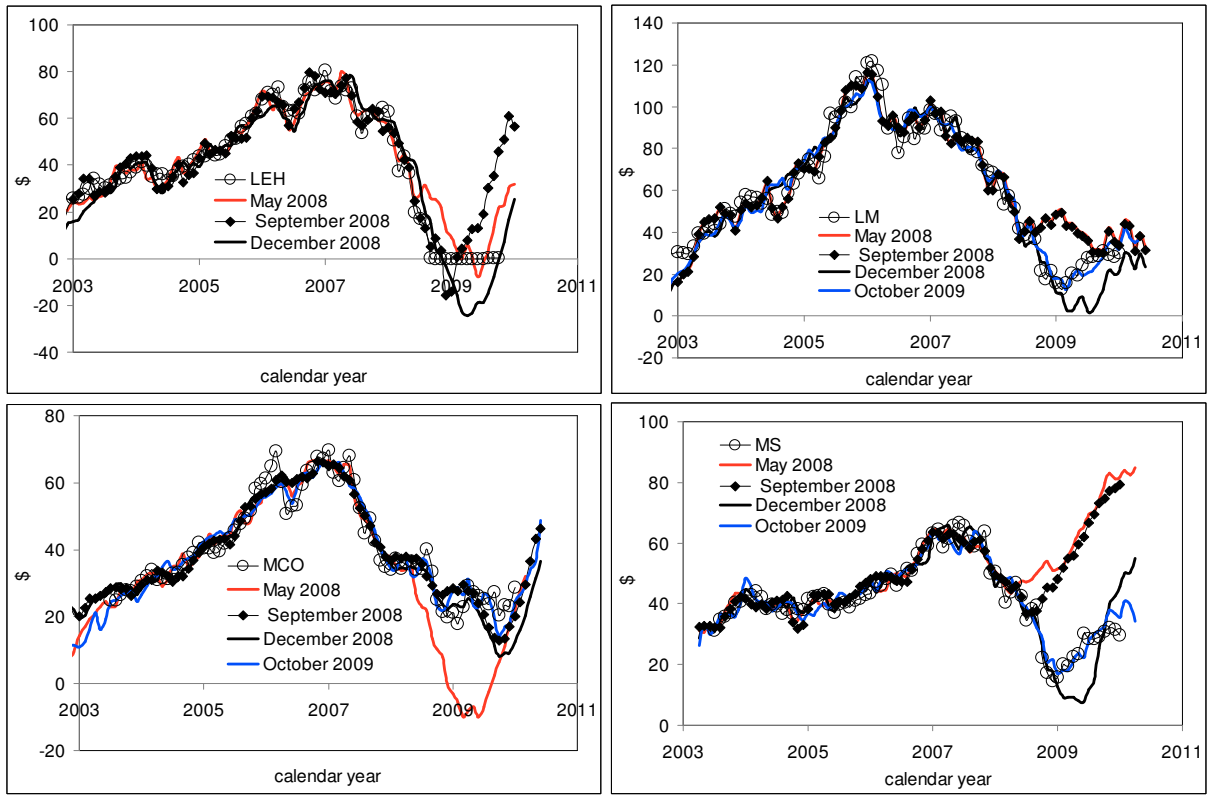

Figure 3. Comparison of stock prices for several financial companies as predicted in May, September and December 2008, and October 2009

There are six companies, all with predicted negative prices but different fate. We have a question on relative merits of the previous bank bailouts - which bank did deserve a bailout and how much would it really cost? The models in Table 4, although they are only tentative ones and should be used with all necessary precautions, might provide a measure of debt size. One can estimate the debt as a product of the number of shares and relevant market price, which was negative for the bailed out and not bailed out companies. Table 5 lists the estimated debts. Lehman Brothers had a much smaller debt than that of Citigroup, CIT and AIG. So, it would have been much easier to bail out LEH from the mathematical point of view. Also, the joint debt of AIG, FRE and FNM is less than $200 billion.

So, we have answered all questions formulated in the beginning of this Section. When having valid pricing models for the companies under consideration, one could foresee all problems before they become serious and select appropriate measures including bailouts. Moreover, taking into account the deterministic evolution of the CPI and linear trends in the CPI differences (Kitov and Kitov, 2008), one could predict major problems long before they happen and avoid most of the 2008/2009 turmoil. For this, financial companies should learn the CPI components defining the evolution of their stocks.

Table 5. Total debt as calculated from negative share prices.

| Company | ## Shares | Share price, $ | Debt, $ |
|---|---|---|---|
| LEH | $6.89 \cdot 10^8$ | -20 | $1.4 \cdot 10^{10}$ |
| C | $1.1 \cdot 10^9$ | -30 | $3.3 \cdot 10^{11}$ |
| CIT | $8.12 \cdot 10^9$ | -20 | $1.6 \cdot 10^{11}$ |
| AIG | $1.34 \cdot 10^8$ | -360 | $1.0 \cdot 10^{11}$ |
| FRE | $6.8 \cdot 10^8$ | -40 | $2.6 \cdot 10^{10}$ |
| FNM | $1.11 \cdot 10^9$ | -50 | $5.5 \cdot 10^{10}$ |

**Discussion**

A deterministic model has been developed for the prediction of stock prices at a horizon of several months. The model links the shares of traded companies to consumer price indices. In this paper, we presented empirical models for financial companies from the S&P 500 list. In May 2008, the model predicted negative share prices in the second half of 2008 for Lehman Brothers, American International Group, Freddie Mac. With known defining CPI components one could predict the approaching bankruptcies. This makes of crucial importance the estimation of correct empirical models, i.e. defining CPIs, for all shares. When reversed, the model also makes it is possible to predict the evolution of various CPI subcategories.

Despite its apparent opposition to the mainstream concepts, the pricing model is deeply rooted in economics: a higher pricing power achieved by a given company should be converted into a faster growth in corresponding consumer price index. This link works excellent for many S&P 500 companies. A further improvement in the model's predictive power is likely possible using advanced methods of statistical and econometrical analysis. However, one should bear in mind that the model will work until its influence on the market is negligible. When a good portion of market participants uses the model it should fail because the market functioning will be disturbed.

Observed and predicted share prices are measured variables and the link between them is likely of a causal character during the studied period. Therefore, the mainstream stock pricing models are, in part, valid – when the evolution of the driving force is random the price is also random, but predictable.

An important possibility arises from our analysis. Using different subsets of the CPI, one can improve our tentative models for the studied companies, and easily obtain similar quantitative relationships for other companies. By extrapolating previously observed trends into the future, one may forecast share prices at various horizons. What likely is more important for a broader investor community, the proposed model also allows predicting the turning points between adjacent trends, when share prices are subject to a substantial decline.

The presented results are preliminary ones and do not pretend to provide an optimal price prediction. A comprehensive investigation with smaller components of the CPI will likely give superior results. So, we recommend refining the model in order to obtain accurate quantitative results for actual investment strategies. All in all, the lagged differences between two CPI components provide a good approximation for the evolution of many stock prices.

One may pose a question: Why did the researches in economics and finances fail to derive the model many years ago? The answer is a scientific one. There were no appropriate data. First, the partition of the headline CPI in hundreds of components is a very new development. Moreover, this process is ongoing and a researcher obtains a more adequate set of defining variables. This brings both higher resolution and reliability. Second, the reliability critically depends on the dynamic range of data. The crisis of 2008 and 2009 has resulted in a dramatic change in both share prices and CPI components. The increased resolution and dynamic range allowed deriving a sound quantitative model. There was no chance to find the link between the share prices and CPI before the data allow. This is a general consideration applicable to all economic and financial models – adequate data must come first (Kitov, 2009a).

Appendix 1. List of seventy CPI components used in the study; in alphabetic order

| Acronym | Description | Acronym | Description |
| --- | --- | --- | --- |
| A | apparel | MAP | men's and boy's apparel |
| AB | alcoholic beverages | MCC | medical care commodities |
| APL | appliances | MCS | medical care services |
| C | CPI | MEAT | meats, poultry, and fish |
| CC | core CPI | MF | motor fuel |
| CE | CPI less energy | MISG | miscelleneous goods |
| CF | CPI less food | MISS | miscellenous services |
| CFSH | CPI less food and shelter | MVI | motor vehicle insurance |
| CFSHE | CPI less food shelter and energy | MVP | motor vehicle parts |
| CM | CPI less medcare | MVR | motor vehicle repaire |
| CO | communication | NC | new cars |
| COMM | commodities | NDUR | nondurables |
| CSH | CPI less shelter | O | other goods and services |
| DIAR | diary products | ORG | other recreation goods |
| DUR | durables | OS | other services |
| E | energy | PC | personal care |
| EC | education and communication | PDRUG | prescription drugs |
| ED | education | PETS | pets and related goods |
| F | food and beverages | R | recreation |
| FB | food less beverages | RENT | rent |
| FISH | fish | RPR | rent primary residence |
| FOOT | footware | RRM | recreational reading materials |
| FOTO | photography | RS | recreation services |
| FRUI | fruits and vegetables | SEFV | food away from home |
| FS | financial services | SERV | services |
| FU | fuels and utilities (housing) | SH | shelter |
| H | housing | SPO | sporting goods (apparel) |
| HFO | household furnishing and operations | T | transportation |
| HO | household operations | TOB | tobacco |
| HOSP | hospital services | TPR | private transportation |
| HS | housekeeping supplies | TPU | public transportation |
| ITR | intracity transportation | TS | transportation services |
| JEW | jewelry and watches | TUIT | tuition |
| LS | legal services | VAA | video and audio |
| M | medical care | WAP | women's and girl's apparel |

Appendix 2. Empirical 2-C models for S&P 500 financial companies

| Company | $b_1$ | $CPI_1$ | $\tau_1$ | $b_2$ | $CPI_2$ | $\tau_2$ | $c$ | $d$ | $\sigma$, $ |
|---|---|---|---|---|---|---|---|---|---|
| AIG | -191.36 | SEFV | 1 | 38.53 | PDRUG | 13 | 727.81 | 21166.78 | 92.31 |
| C | 2.94 | HO | 5 | -8.26 | SEFV | 2 | 36.70 | 1048.90 | 2.53 |
| CB | -1.27 | F | 3 | -0.41 | O | -3 | 16.53 | 313.52 | 1.70 |
| CINF | -0.06 | TOB | -3 | -2.79 | SEFV | -4 | 21.53 | 486.92 | 1.41 |
| CME | -38.62 | PETS | 0 | -30.86 | AB | 10 | 458.48 | 8475.35 | 41.43 |
| EQR | -3.16 | SEFV | 3 | 1.12 | PDRUG | 4 | 11.48 | 187.90 | 2.14 |
| FHN | 1.52 | MCC | 8 | -4.10 | SEFV | -5 | 14.30 | 340.70 | 1.90 |
| FII | 1.49 | HO | 12 | -1.78 | PETS | 2 | 4.02 | 38.95 | 2.01 |
| HCBK | 0.03 | E | 1 | 0.47 | MISS | -2 | -4.43 | -113.67 | 0.83 |
| HCP | 1.31 | MCC | 5 | 0.30 | FS | -1 | -8.35 | -365.78 | 2.10 |
| HES | 3.37 | CSH | -2 | 3.81 | MISS | -5 | -47.75 | -1480.90 | 4.87 |
| HIG | 0.53 | TPU | 12 | -8.35 | PETS | 1 | 44.41 | 741.76 | 4.79 |
| HST | -1.30 | FB | 4 | -1.39 | RPR | 11 | 18.91 | 451.01 | 1.12 |
| IVZ | 2.44 | HO | 11 | -1.52 | PETS | 3 | -1.04 | -99.66 | 1.79 |
| JNS | -2.99 | PETS | 2 | 3.14 | RPR | 7 | -4.93 | -257.86 | 2.63 |
| KEY | -0.36 | DIAR | 9 | -3.92 | RPR | 11 | 28.68 | 763.73 | 1.59 |
| KIM | 3.15 | RS | -3 | -5.14 | SEFV | 2 | 25.07 | 454.58 | 2.17 |
| LNC | -4.59 | F | 5 | -2.35 | TS | 3 | 41.28 | 1212.45 | 3.80 |
| LUK | 0.65 | TPR | -2 | -1.42 | MVI | 3 | 7.66 | 332.12 | 3.13 |
| MCO | -0.95 | MEAT | 4 | -8.58 | RPR | 9 | 69.27 | 1664.39 | 3.90 |
| MET | 0.36 | TPU | 13 | -4.41 | PETS | 2 | 26.48 | 364.93 | 2.88 |
| MI | -2.54 | SEFV | 4 | -2.44 | RPR | 13 | 32.35 | 850.83 | 1.64 |
| MS | 5.27 | HO | 8 | -9.75 | SEFV | 2 | 39.55 | 1031.57 | 3.72 |
| MTB | -3.96 | FB | 3 | -4.65 | RPR | 11 | 57.08 | 1510.23 | 4.87 |
| NTRS | -3.02 | PETS | 2 | 3.66 | RPR | 5 | -5.16 | -340.09 | 3.94 |
| PBCT | 0.65 | MCC | 7 | -0.75 | MVP | 13 | 0.12 | -82.64 | 0.77 |
| PCL | -0.81 | MCC | -1 | 0.57 | FS | -2 | 5.27 | 79.01 | 2.02 |
| PFG | -3.01 | PETS | -2 | 0.97 | FS | -2 | 14.95 | 85.06 | 3.55 |
| PGR | 0.13 | FU | -2 | -1.74 | RPR | 2 | 11.47 | 309.11 | 1.29 |
| PLD | -3.09 | PETS | -1 | 1.17 | FS | -2 | 14.18 | 42.98 | 3.44 |
| PRU | -8.13 | PETS | 2 | 0.18 | TOB | 0 | 45.25 | 723.38 | 5.44 |
| RF | -1.29 | F | -3 | -2.04 | FB | 7 | 18.09 | 554.44 | 1.48 |
| SLM | 1.91 | PETS | 13 | -9.31 | RPR | 12 | 53.36 | 1490.40 | 4.12 |
| SPG | -5.78 | F | 3 | 0.98 | FS | -1 | 38.82 | 693.54 | 5.36 |
| STI | -5.70 | FB | 4 | -0.16 | TOB | 5 | 37.05 | 1010.63 | 3.65 |
| STT | 5.12 | HO | 11 | -4.61 | PETS | 4 | 3.78 | -50.41 | 5.02 |
| TMK | -8.45 | SEFV | 3 | 0.77 | HOSP | 2 | 36.15 | 1140.20 | 3.24 |
| TROW | -1.54 | FB | 3 | -2.09 | TS | 5 | 26.41 | 644.30 | 3.51 |
| UNM | -0.17 | FU | 6 | -0.91 | PETS | 2 | 8.82 | 110.67 | 1.56 |
| USB | -1.10 | FB | 4 | 0.32 | FS | 0 | 6.26 | 117.85 | 1.67 |
| VTR | 0.45 | FS | -2 | -1.60 | MISS | 8 | 18.93 | 276.81 | 2.41 |
| WFC | -1.15 | O | -3 | 0.09 | TOB | 0 | 10.72 | 281.54 | 2.06 |
| XL | -13.65 | RPR | 13 | 6.35 | MVR | 13 | 45.24 | 1413.75 | 4.37 |
| ZION | -2.19 | F | 4 | -8.09 | RPR | 13 | 67.14 | 1812.98 | 3.21 |

Appendix 3. Observed and predicted stock prices of S&P 500 financial companies

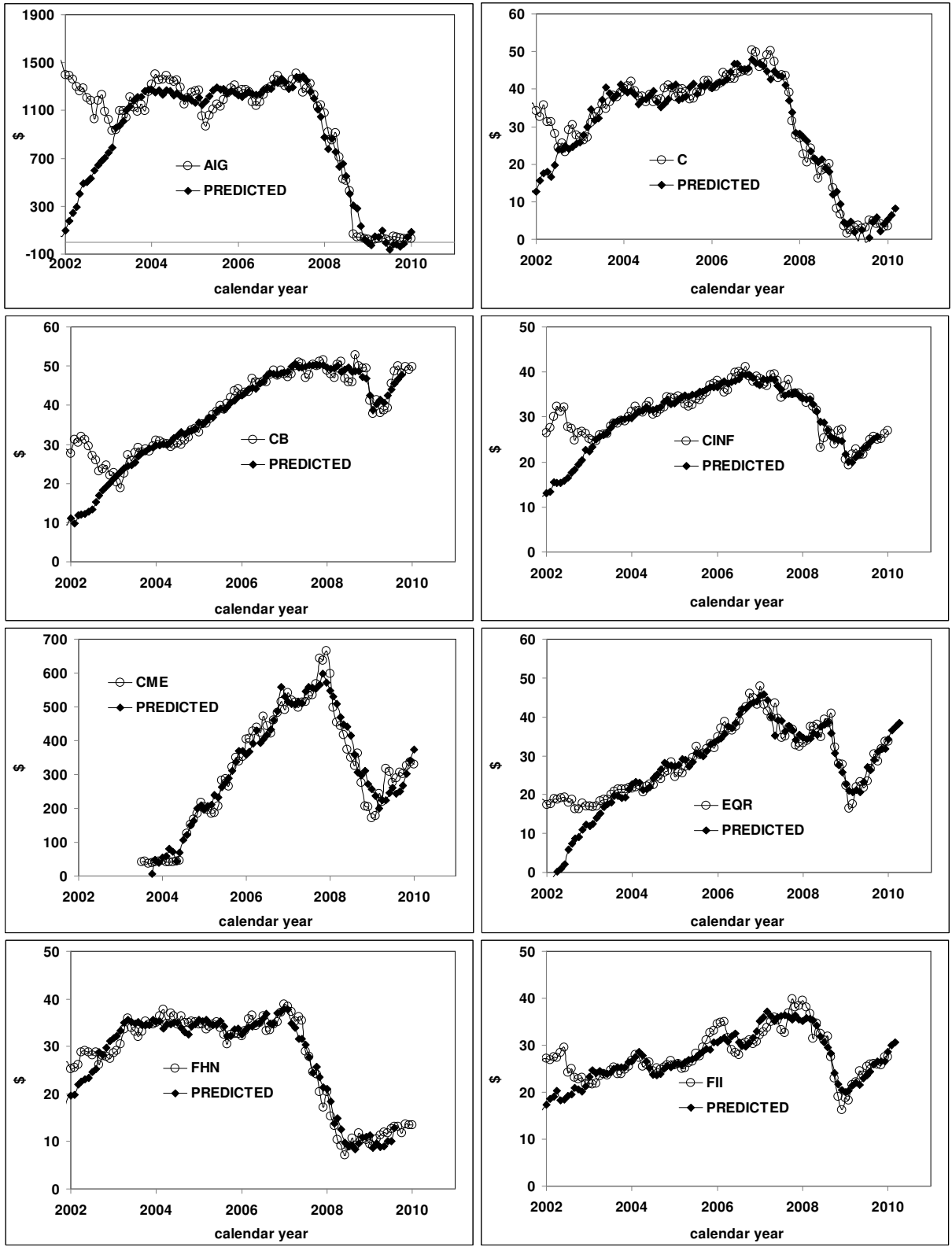

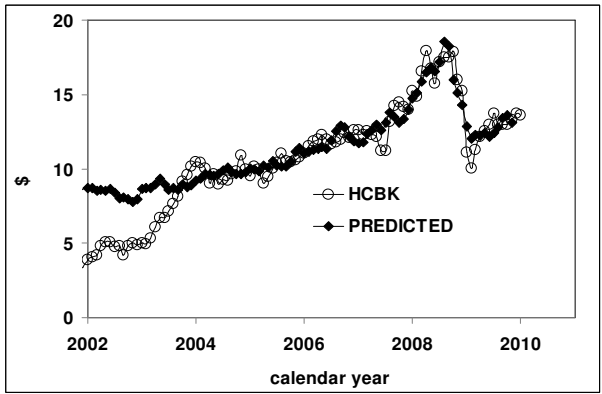
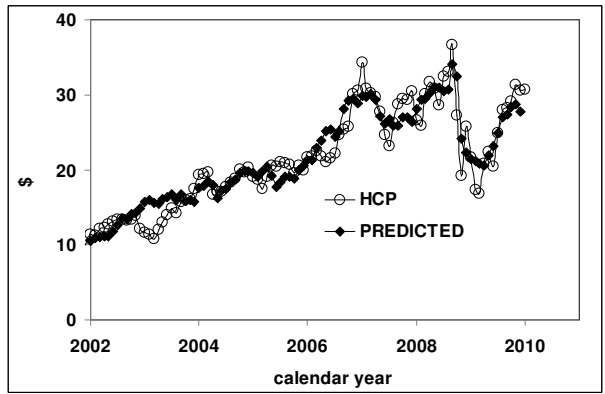
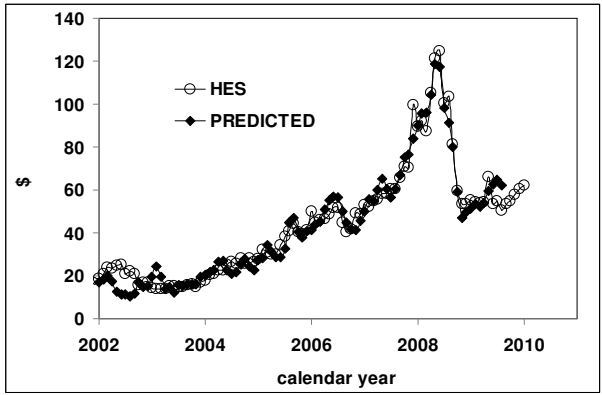
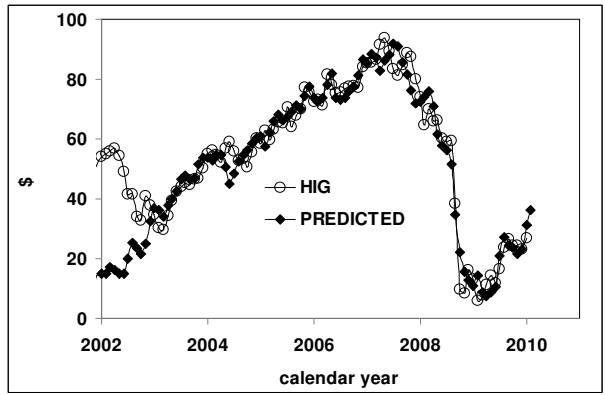
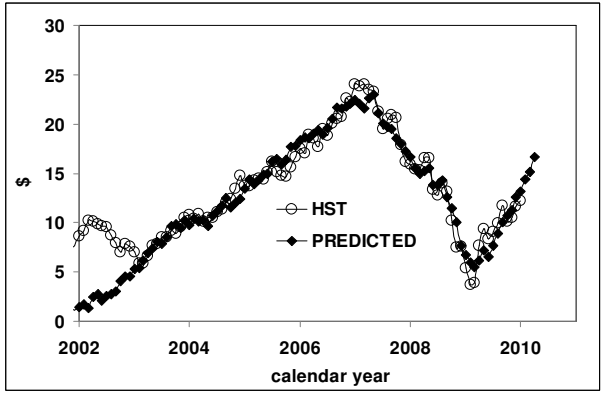
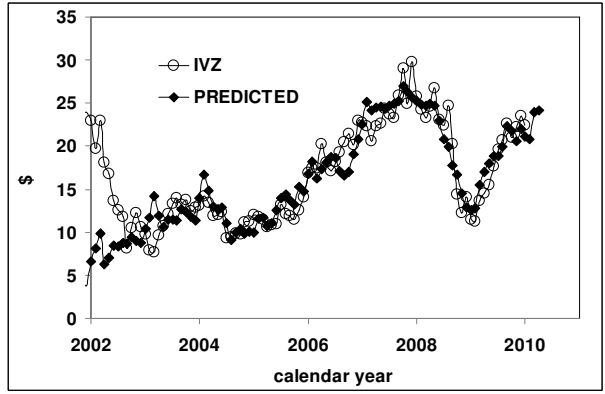
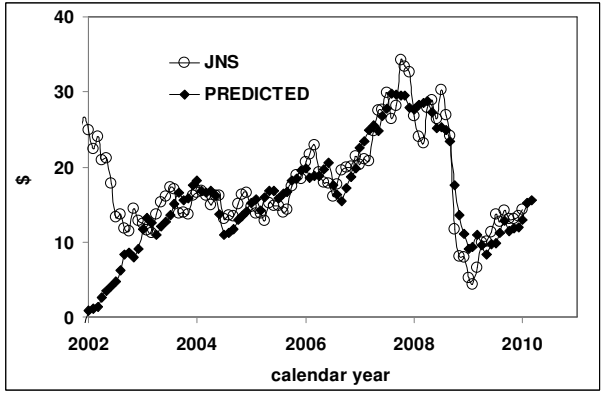
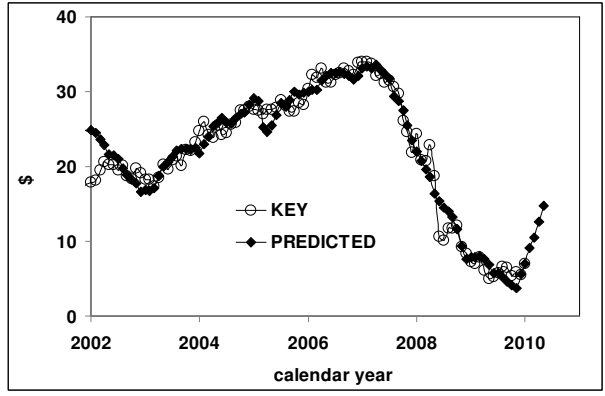

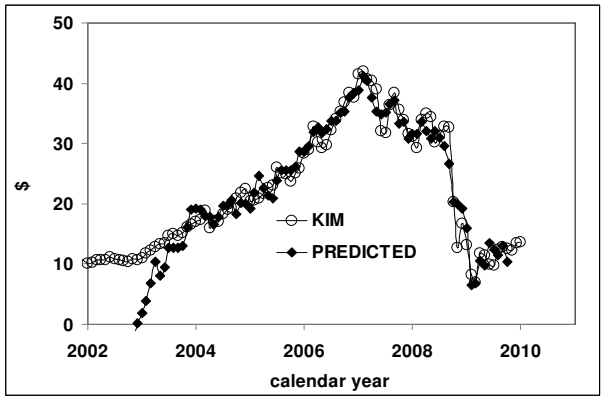
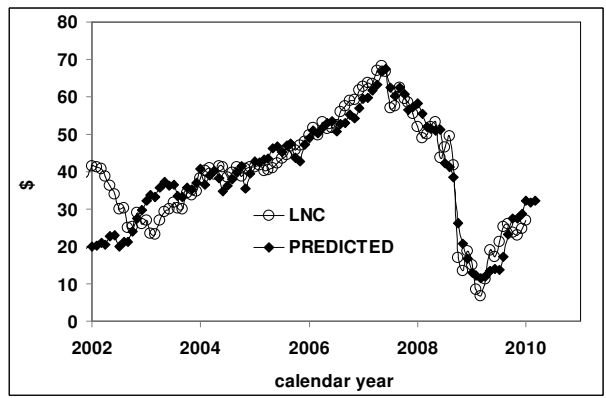
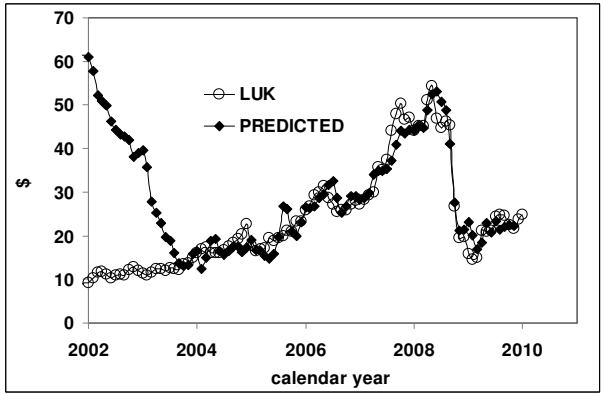
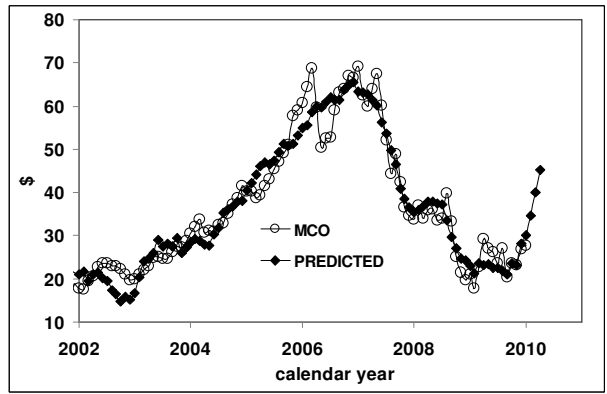
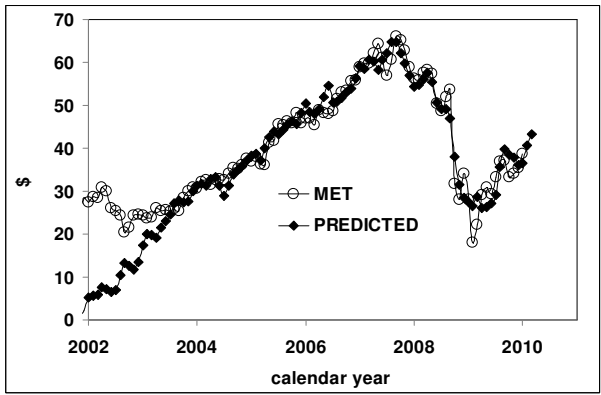
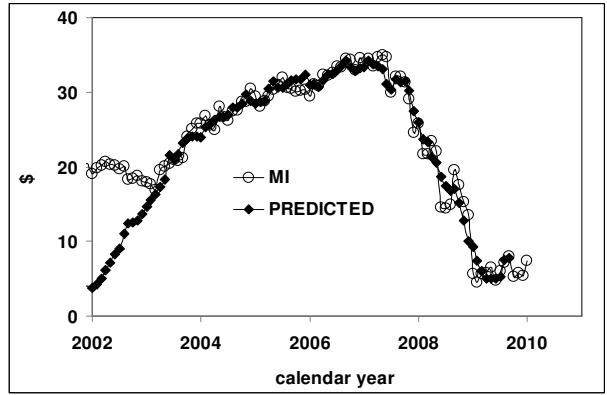
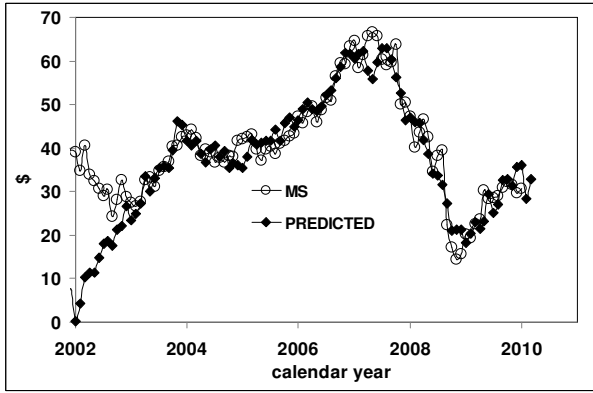
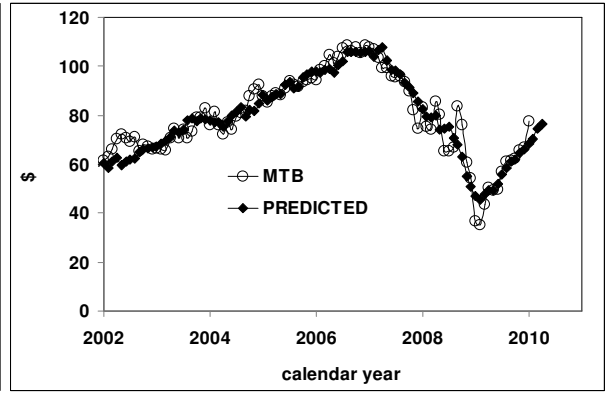

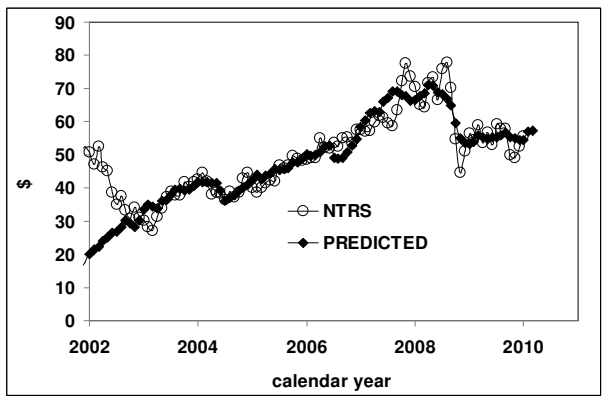
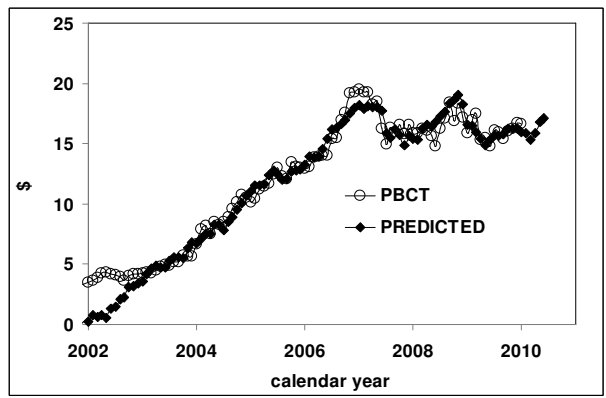
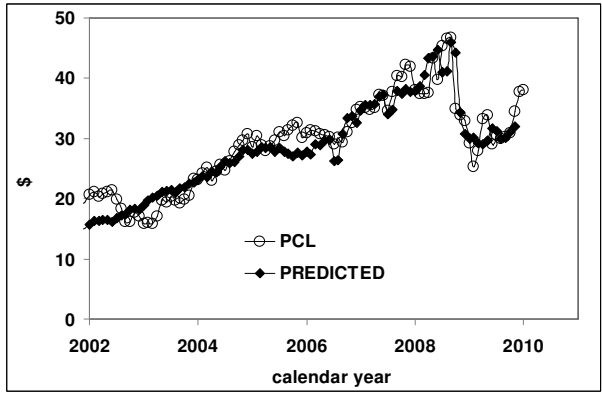
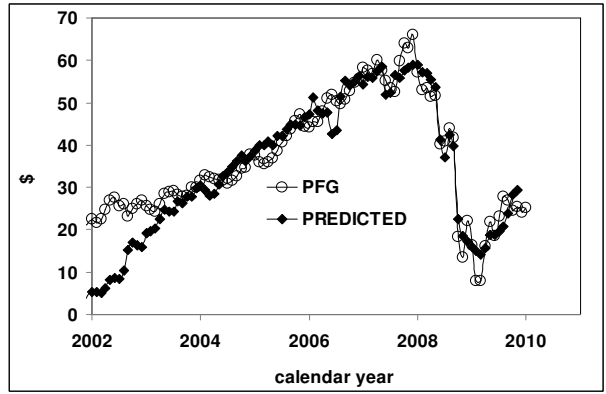
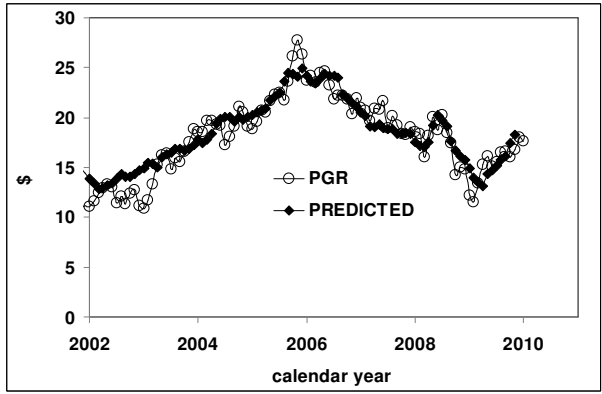
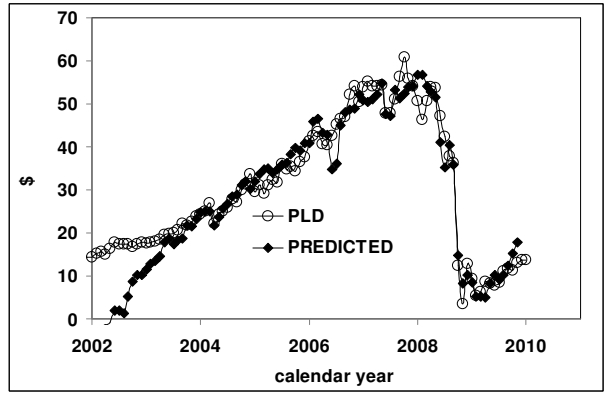
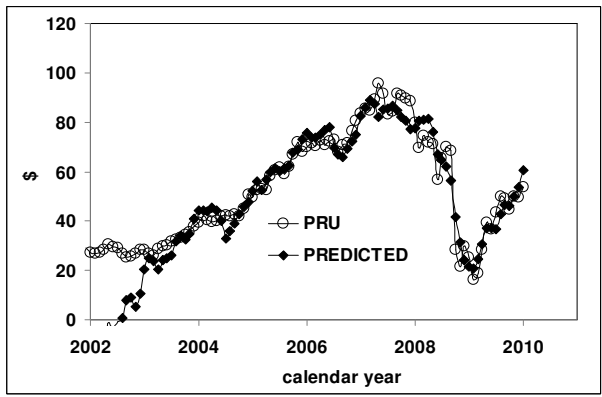
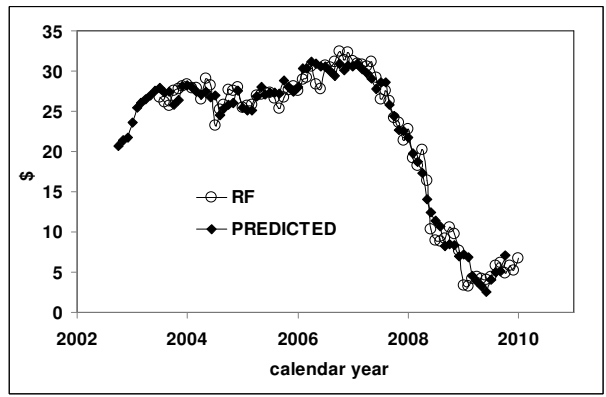

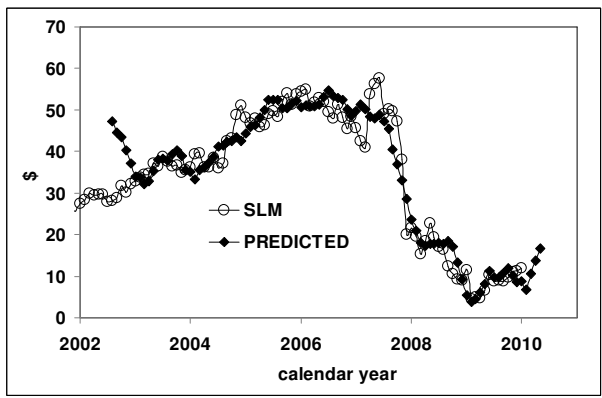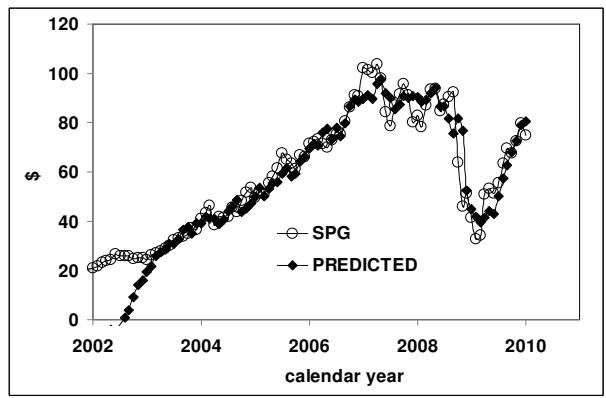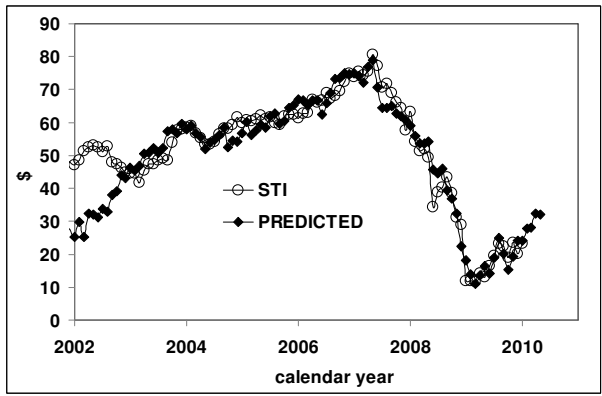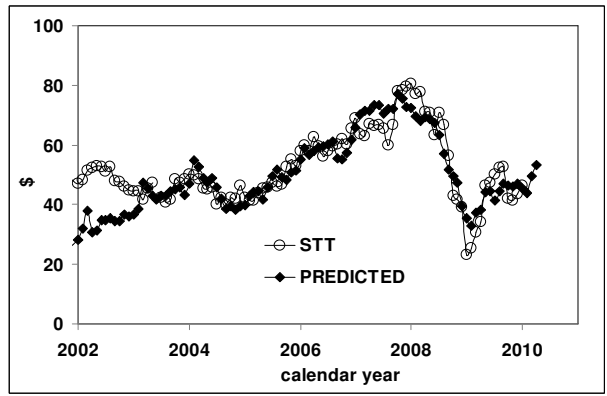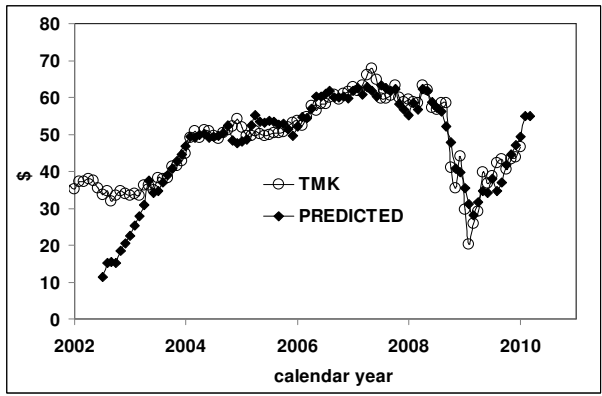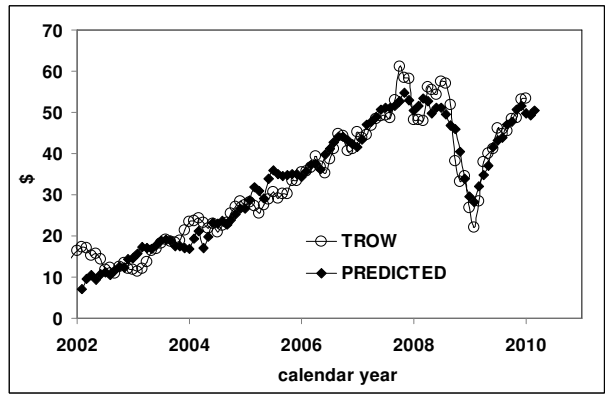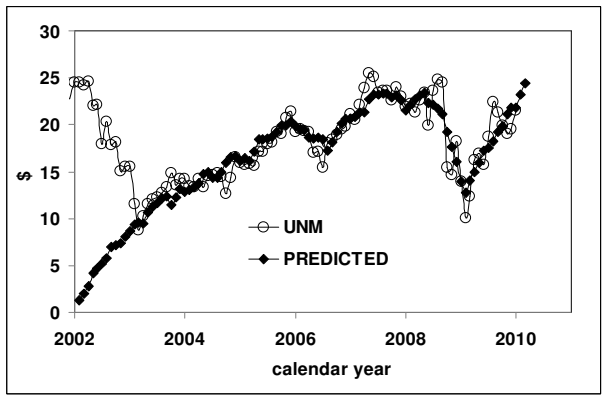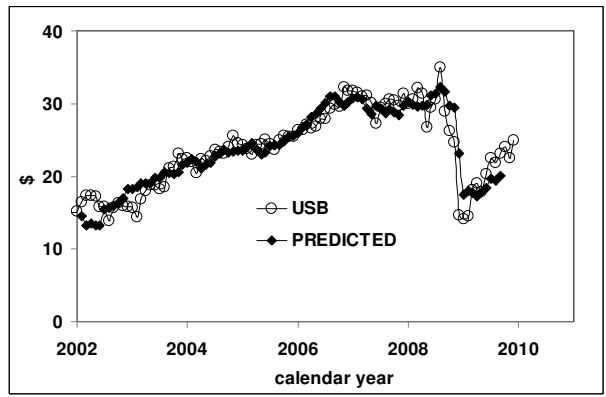

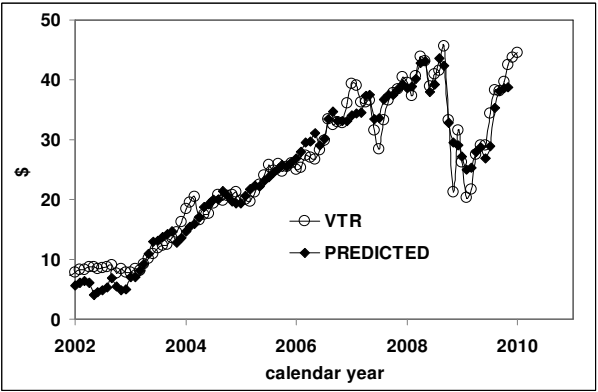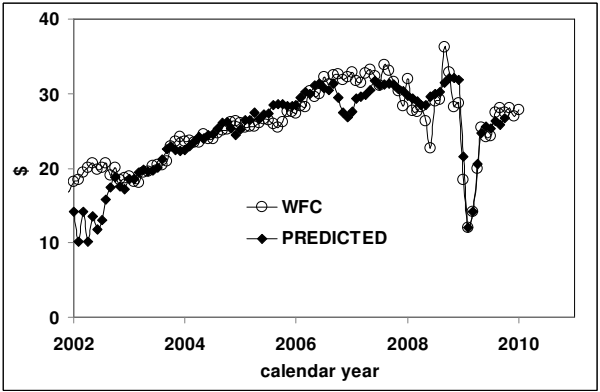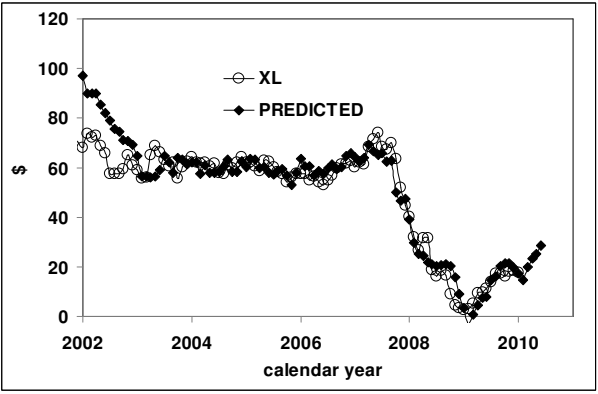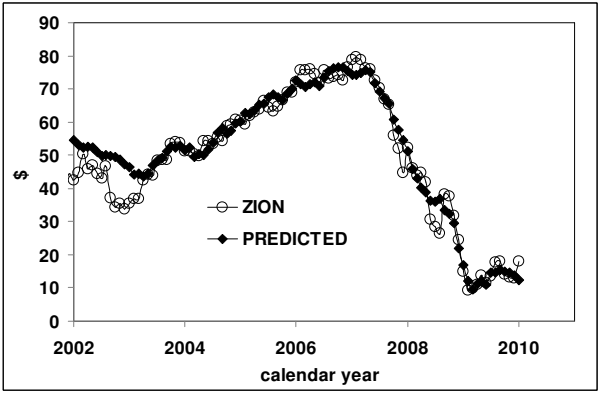

Appendix 4. Empirical models, as of May 2008.

| Company | $b_1$ | $CPI_1$ | $\tau_1$ | $b_2$ | $CPI_2$ | $\tau_2$ | $c$ | $d$ |
|---|---|---|---|---|---|---|---|---|
| CB | -0.39 | FISH | 8 | -2.48 | RS | 13 | 17.32 | 335.17 |
| CINF | -2.09 | F | 8 | -3.41 | PC | 1 | 27.78 | 906.09 |
| CME | 6.79 | TOB | 6 | -169.02 | SEFV | 6 | 1014.46 | 23534.41 |
| EQR | 3.33 | MVP | 5 | -5.53 | RPR | 6 | 31.91 | 670.54 |
| FHN | -6.62 | **SEFV** | 4 | 1.02 | PDRUG | 8 | 25.99 | 810.51 |
| FII | 2.27 | DUR | 10 | -1.54 | HS | 1 | 11.38 | -48.93 |
| HCBK | 0.07 | TOB | 13 | -0.25 | PDRUG | 9 | 3.12 | 45.50 |
| HCP | 4.91 | R | 5 | 0.94 | OS | 3 | -9.23 | -713.62 |
| HES | -7.91 | HO | 5 | 17.05 | SEFV | 9 | -44.26 | -1925.87 |
| HIG | -5.74 | RPR | 7 | 2.65 | PDRUG | 13 | 17.33 | 307.40 |
| IVZ | -0.46 | MEAT | 7 | -0.39 | MVI | 4 | 8.04 | 181.92 |
| JNS | 1.14 | DUR | 11 | -2.97 | FOTO | 5 | -4.86 | 183.10 |
| KEY | -3.14 | **RPR** | 6 | 0.98 | PDRUG | 11 | 11.84 | 304.61 |
| KIM | 0.98 | PDRUG | 6 | 2.00 | FOTO | 11 | 0.13 | -496.62 |
| LNC | 3.12 | MVP | 4 | -10.95 | SEFV | 4 | 61.51 | 1455.18 |
| LUK | -4.26 | VAA | 8 | 1.74 | OS | 13 | -5.82 | 46.26 |
| MCO | 3.69 | DUR | 9 | -9.51 | **RPR** | 8 | 75.41 | 1232.98 |
| MET | -1.56 | SH | 11 | 1.13 | PDRUG | 12 | 4.74 | -22.58 |
| MI | -0.86 | FB | 3 | -5.19 | **SEFV** | 4 | 37.48 | 979.40 |
| MTB | -7.20 | **RPR** | 11 | -3.57 | MISS | 12 | 87.49 | 2180.14 |
| NTRS | -1.10 | MEAT | 8 | 0.45 | DIAR | 2 | 11.12 | 102.13 |
| PBCT | -0.73 | O | 3 | 0.38 | PDRUG | 7 | 4.68 | 81.62 |
| PCL | -0.22 | FU | 6 | -0.16 | TPU | 0 | 7.81 | 59.81 |
| PFG | 1.57 | APL | 3 | 6.50 | FOTO | 13 | 25.34 | -840.96 |
| PGR | 0.40 | MVI | 9 | 1.45 | RRM | 12 | -5.52 | -368.27 |
| PLD | -2.16 | FB | 4 | -1.02 | **FS** | 13 | 26.75 | 547.08 |
| PRU | 2.00 | PDRUG | 13 | -1.65 | HOSP | 4 | 31.56 | -71.92 |
| RF | -3.31 | RPR | 13 | -1.89 | RRM | 3 | 25.57 | 967.64 |
| SLM | 4.67 | VAA | 12 | -6.52 | **RPR** | 13 | 41.17 | 706.86 |
| SPG | 15.04 | R | 5 | -5.63 | PETS | 3 | 24.43 | -1021.99 |
| STI | -8.07 | SEFV | 0 | 1.57 | PDRUG | 11 | 35.94 | 893.58 |
| STT | -3.79 | SERV | 11 | 5.71 | **HO** | 11 | 8.40 | 131.84 |
| TMK | 1.22 | MCC | 8 | 0.59 | ITR | 8 | -9.63 | -348.07 |
| TROW | 1.54 | EC | 0 | 0.75 | PDRUG | 13 | -4.58 | -370.49 |
| UNM | -1.09 | FB | 4 | -0.84 | MCC | 4 | 14.36 | 374.29 |
| USB | -0.04 | E | 1 | -1.23 | MISS | 12 | 15.68 | 307.26 |
| VTR | 1.04 | MCC | 4 | 0.85 | **FS** | 10 | -7.35 | -430.76 |
| WFC | -1.62 | F | 13 | -1.38 | FB | 0 | 18.21 | 491.21 |
| XL | -2.86 | O | 12 | 2.95 | PDRUG | 11 | -14.86 | 19.28 |
| ZION | -3.22 | AB | 3 | -7.99 | **RPR** | 13 | 70.08 | 1989.05 |